\documentclass{ar-1col}
\usepackage{epsfig}
\usepackage{color}
\usepackage{natbib}
\usepackage{subcaption}
\usepackage{amsmath, amssymb, amstext}
\usepackage{mathtools}

\usepackage{todonotes}

\jname{Annu. Rev. Astron. Astrophys.}
\jvol{AA}
\jyear{2018}
\doi{10.1146/(please add article doi)}

\def\mgal{M_{\ast}}
\def\mhalo{M_h}
\def\slogm{\sigma_{\log M_\ast}}
\def\msun{{\rm M}_\odot}
\def\msol{{\rm M}_\odot}
\def\meanmgal{\langle \mgal\rangle}

\def\shmrr{{\rm SHMR_{red}}}
\def\shmrb{{\rm SHMR_{blue}}}
\def\mred{M_{\ast{\rm red}}}
\def\mblue{M_{\ast{\rm blue}}}
\def\zhalf{z_{M/2}}
\def\mnl{M_{\rm \scriptscriptstyle NL}}

\def\fq{f_q}

\def\fsat{f_{\rm sat}}
\def\DS{\Delta\Sigma(R_p)}
\def\vmax{V_{\rm max}}

\begin{document}

\markboth{Wechsler \& Tinker}{The Galaxy--Halo Connection}

\title{The Connection between Galaxies and their Dark Matter Halos}

\author{Risa H. Wechsler$^1$ and Jeremy L. Tinker$^2$
\affil{$^1$Kavli Institute for Particle Astrophysics and Cosmology and Department of Physics, Stanford University, Stanford, CA, USA, 94305; Department of Particle Physics \& Astrophysics, SLAC National Accelerator Laboratory, Menlo Park, CA 94025, email: rwechsler@stanford.edu}
\affil{$^2$Center for Cosmological Physics, Department of Physics, New York University, New York, NY, USA, 10003; email: jeremy.tinker@nyu.edu}}

\begin{abstract}
In our modern understanding of galaxy formation, every galaxy forms within a dark matter halo. The formation and growth of galaxies over time is connected to the growth of the halos in which they form.  The advent of large galaxy surveys as well as high-resolution cosmological simulations has provided a new window into the statistical relationship between galaxies and halos and its evolution.  
Here we define this galaxy--halo connection as the multi-variate distribution of galaxy and halo properties that can be derived from observations and simulations. This connection provides a key test of physical galaxy formation models; it also plays an essential role in constraints of cosmological models using galaxy surveys and in elucidating the properties of dark matter using galaxies.  We review techniques for inferring the galaxy--halo connection and the insights that have arisen from these approaches.  Some things we have learned are that galaxy formation efficiency is a strong function of halo mass; at its peak in halos around a pivot halo mass of $10^{12}~ \msun$, less than 20\% of the available baryons have turned into stars by the present day; the intrinsic scatter in galaxy stellar mass is small, less than 0.2 dex at a given halo mass above this pivot mass; below this pivot mass galaxy stellar mass is a strong function of halo mass; the majority of stars over cosmic time were formed in a narrow region around this pivot mass. We also highlight key open questions about how galaxies and halos are connected, including understanding the correlations with secondary properties and the connection of these properties to galaxy clustering.

\end{abstract}

\maketitle

\tableofcontents

\section{INTRODUCTION}
In modern cosmological models, $\sim 5/6$ of the mass in the Universe is made of dark matter \citep{planck_2016}. This dark matter forms the skeleton on which galaxies form, evolve, and merge.  In the context of this model, which is now well established from a wide range of observations, fluctuations in the matter distribution were created in the first fraction of a second during an inflationary period.  Gravitational instability grew these fluctuations over time.  Gas and dark matter were initially well mixed; as the universe evolved, gas was able to dissipate and fell to the centers of dark matter halos.  For large enough dark matter halos, gas was able to cool, form stars, and form a protogalaxy.  The power spectrum of matter indicates that small objects should form first, and halos should grow and merge over time.  Galaxies within these halos then continue to form stars ({\em in situ}) as well as to grow through merging ({\em ex situ}), because their dark matter halos merge.  Energetic processes within galaxies impact their surroundings after they form, as various kinds of feedback, which influences future gas accretion and star formation.  

In this context, clearly the growth, internal properties, and spatial distribution of galaxies are likely to be closely connected to the growth, internal properties, and spatial distribution of dark matter halos.  Very simply, the luminous matter in the Universe is arranged in galaxies, and in a cold dark matter model, the dark matter in the Universe is arranged in dark matter halos. The physical and statistical connection between them is the focus of this review. 
We denote this the {\em galaxy--halo connection}, which in detail can refer to the full multivariate distribution of properties of halos and the galaxies that form within them.  Elucidating this connection is a stepping stone to answering several of the largest questions in astrophysics and cosmology today.  These include the following: 
\begin{itemize} 
\item {\bf Understanding the physics of galaxy formation}: How does gas cool in galaxies, how do stars form, and what determines the dominant feedback processes? How can we best infer physics from available observations including the spatial distribution of galaxies as a function of their properties?  Statistical constraints on the galaxy--halo connection can effectively synthesize diverse datasets and provide essential input into these questions.
\item {\bf Inferring cosmological parameters}: A new generation of deep, wide-area imaging and spectroscopic surveys has substantial constraining power on cosmological models, including the power to distinguish between a cosmological constant, dark energy, and modified gravity; to measure the mass of the neutrino; and to constrain the inflationary potential.  To make full use of these measurements on the widest range of scales, we need to understand how to robustly marginalize over the uncertainties in the galaxy--halo connection.  
\item {\bf Probing the properties and distribution of dark matter}: How can we infer the evolution of the matter distribution and the properties of dark matter halos using surveys of galaxies?  Measurements of galaxies and of their spatial distribution give us the potential to map out the distribution of dark matter as well as to distinguish between models of dark matter.  These observations are generally sensitive to combinations of both the dark matter particle and the galaxy--halo connection.  
\end{itemize}

The galaxy--halo connection as a concept started with the earliest understanding of modern galaxy formation within the framework of cold dark matter (CDM) models, as did the understanding that the spatial distribution of galaxies can lead to insight into their formation properties.  For example, \citet{Peebles1980} discussed the two-point statistics of galaxies, and early work recognized that massive galaxies and clusters should have different clustering properties than average galaxies \citep{Davis83, Bahcall83, Klypin83, Kaiser84}, and that measuring these clustering properties could provide information about the masses of the dark matter halos that they lived in, because of the strong dependence of halo clustering on halo mass \citep{bbks,mowhite:96}. It also was recognized early on that this relationship could be complex and scale dependent \citep[e.g.][]{Klypin96,Jenkins98}.  

However, it was not until the late 1990s that cosmological simulations were able to resolve the substructures within larger dark matter halos.   At about the same time, the first large galaxy surveys were beginning.  The APM survey \citep{Baugh1996} was the first to measure the galaxy correlation function for a large sample of galaxies.  \citet{kravtsov_klypin:99} and \citet{colin_etal:99} were able to resolve substructures in simulations of cosmological volumes, and they measured approximately power-law correlation functions that were consistent with measurements from APM.  This field was then revolutionized with the Two-degree Field Galaxy Redshift Survey \citep[2dFGRS;][]{colless_etal:01} and Sloan Digital Sky Survey \citep[SDSS;][]{ york_etal:00}. For the first time, these surveys were able to measure the spatial clustering properties of large samples of galaxies, which allowed for the separation into their physical properties such as luminosity and color or stellar mass and star formation rate (SFR).  Pioneering detections of galaxies at high redshift \citep{adelberger_etal:98} also enabled the first studies of galaxy clustering at these epochs.

These two joint revolutions, {\em (a)} the advent of numerical simulations that could resolve the dark matter structures and substructures hosting galaxies, over volumes large enough to measure their spatial clustering properties and {\em(b)} the advent of large galaxy surveys, that could identify large samples of galaxies and measure their spatial clustering, including over a range of redshifts, have led to a new set of approaches to statistically connect these two distributions and infer the connection between galaxies and halos, that has flourished over the last $\sim$ 15 years.\footnote{We note that prior to 1999, the only use of the phrase ``the galaxy--halo connection'' in the literature was to describe the Milky Way Galaxy and its stellar halo; see e.g. \citet{1996ASPC...92..474V}
.} The primary focus of this review is on the inference of this statistical connection between galaxies and halos enabled by these two advances. We will highlight {\em (a)} theoretical approaches to the problem {\em (b)} the primary insights that we have gained from studying the galaxy--halo connection and {\em (c)} outstanding issues.  We note that the development of the galaxy--halo connection is connected to development of the ``halo model" \citep{ma_fry:00,Peacock2000,Seljak2000,cooray_sheth:02}, a method for analytically calculating the non-linear clustering of dark matter, using the properties of dark matter halos (including their abundance and spatial clustering) as the basic unit.  The halo model can be combined with models of the galaxy--halo connection to predict galaxy clustering.  

Elucidating the statistical connection between galaxies and halos relies on another major advance of the last two decades: the establishment of a standard cosmological model,  $\Lambda$CDM, in which the universe consists of 5\% baryonic matter, 25\% dark matter, and 70\% dark energy.  The parameters of this model are now known to high precision \citep{betoule_etal:14,planck_2016, des_2017,alam_etal:17}; this allows robust predictions, using numerical simulations, for the growth of structure and the formation and evolution of dark matter halos.  Although this review relies heavily on basic predictions of the $\Lambda$CDM model and the properties of dark matter halos, these are well-reviewed elsewhere; see for example \citet{frenk_white:12} and \citet{primack:12}.  The cosmological simulations that form the basis of many of the predictions described here were reviewed by \citet{kuhlen_etal:12}. The current status of physical models of galaxy formation was reviewed in \citet{somerville_dave:15} and outstanding theoretical challenges were reviewed by \cite{naab_ostriker:17};  the formation of galaxy clusters was reviewed by \citet{kravtsov_borgani:12}.  Modern cosmological probes with galaxy surveys were reviewed in \citet{weinberg_etal:13}; here we focus on cosmological studies that require an understanding of the galaxy--halo connection.  The interplay between the galaxy--halo connection and models of dark matter on small scales was well reviewed by \citet{bullock_boylankolchin:17}, so we only briefly touch on these issues here.

We introduce the methods of modeling and predicting the galaxy--halo connection in \S \ref{sec:models}.  In \S \ref{sec:observations}, we review the primary observational handles on the galaxy--halo connection.  In \S \ref{sec:complications}, we discuss complications to the simplest modeling approaches.  In \S \ref{sec:constraints} we discuss current observational constraints on the mean and scatter of the relationship between galaxy mass and halo mass; \S \ref{sec:constraints_secondary} expands this discussion to other aspects of the galaxy--halo connection.  We review the primary applications of the galaxy--halo connection in \S \ref{sec:applications}, including understanding galaxy formation physics, constraining cosmological parameters, and mapping and understanding the physics of dark matter.  In \S \ref{sec:status} we summarize the key aspects of the galaxy--halo connection that have been understood from the last decade of studies, assess the outlook for such studies over the next decade, including how they will be influenced by upcoming surveys,  and highlight outstanding questions for future work.
\section{MODELS OF THE GALAXY--HALO CONNECTION}
\label{sec:models}

\begin{figure}[h]
\includegraphics[width=5in]{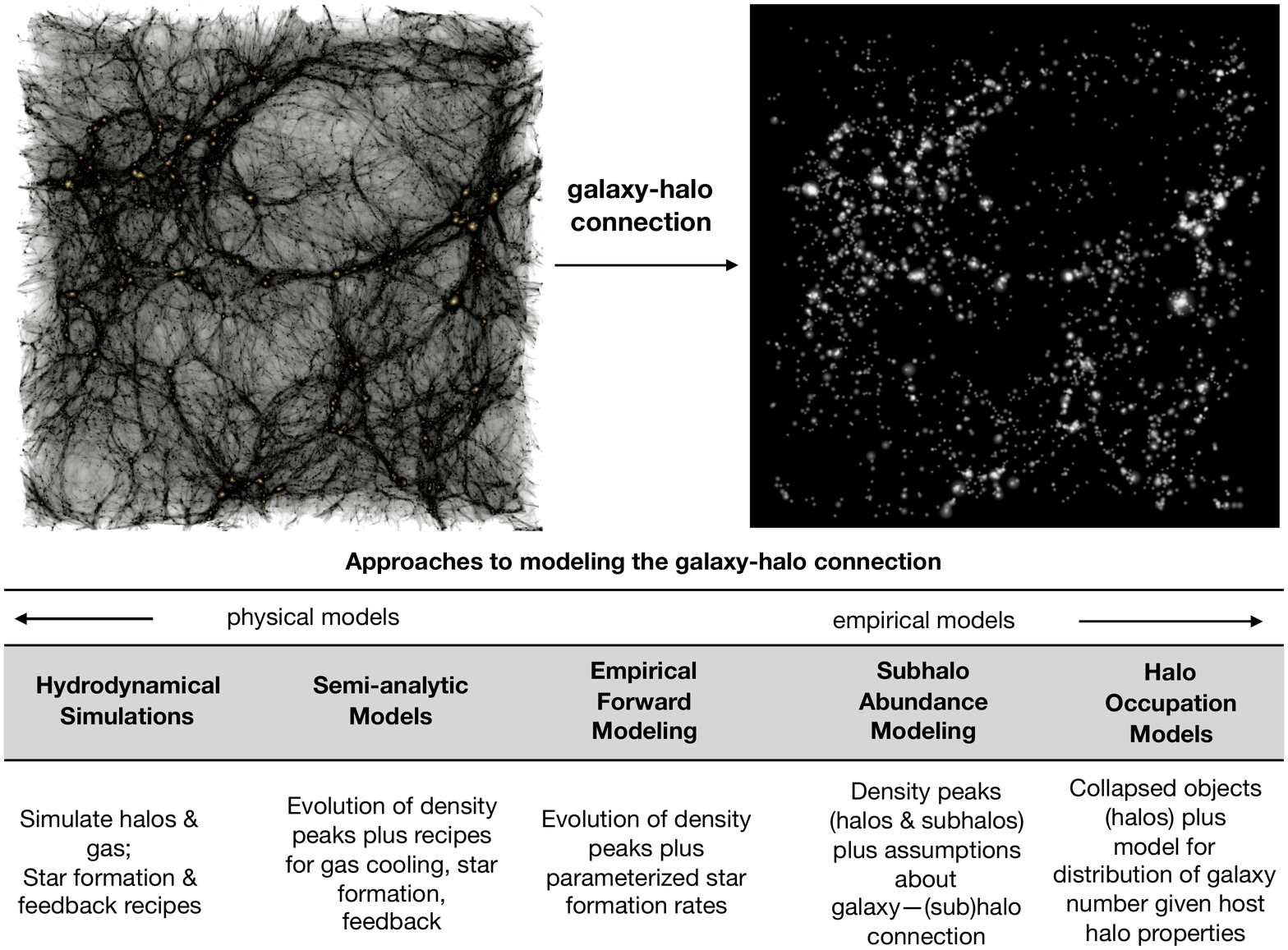}
\caption{\footnotesize{Modeling approaches to the galaxy--halo connection.  Top panel shows the dark matter distribution in a 90$\times$90$\times$30 Mpc $h^{-1}$ slice of a cosmological simulation {\em (Left)}, compared to the galaxy distribution using an abundance matching model, tuned to match galaxy clustering properties of an observed sample ({\em Right}).  The grid highlights the key assumptions of various models for the galaxy--halo connection.  The models are listed on a continuum from left to right ranging from more physical and predictive (making more assumptions from direct simulation or physical prescriptions) to more empirical (more flexible parameterizations, constrained directly from data).}}
\label{models}
\end{figure}

In this review we distinguish between two basic approaches to modeling the galaxy--halo connection, {\em empirical modeling}, which uses data to constrain a specific set of parameters describing the connection at a given epoch or as a function of time, and {\em physical modeling}, which either directly simulates or parameterizes the physics of galaxy formation such as gas cooling, star formation, and feedback.  A schematic summary of these approaches to the galaxy--halo connection is given in \textbf{Figure \ref{models}}, which gives an example of the galaxy and dark matter distributions for one such model, and outlines the key elements of various approaches.  We note that in practice these modeling approaches are more of a continuum: as one moves to the right in this figure, one is assuming less physics directly from the model itself, and has more flexibility to constrain the unknown aspects of the galaxy--halo connection directly with data, but the models are also less predictive and less directly connected to the physical prescriptions.  In general, approaches towards the right are also significantly less expensive computationally than the more physical approaches.

We begin in \S \ref{sec:halo} by reviewing the concept of a dark matter halo.  We then review current approaches to empirical modeling of the galaxy--halo connection in \S \ref{sec:empirical}, including abundance matching, the halo occupation distribution and conditional luminosity function, and models which connect galaxies over time to their histories.  In \S \ref{sec:physical}, we review approaches to physical modeling of galaxy formation, including hydrodynamical simulations and semi-analytic modeling, highlighting areas of synergy with empirical approaches.

\subsection{Preliminaries: What is a halo?}
\label{sec:halo}
In the modern theory of cosmological structure formation, dark matter halos are the basic unit into which matter collapses.  Schematically, halos can be thought of as gravitationally bound regions of matter that have decoupled from the Hubble expansion and collapsed.  In numerical simulations they are generally defined with masses and radii specified by a given overdensity: $M_{\rm vir} = \tfrac{4\pi}{3}R_{\rm vir}^3\Delta\rho_m$.  The definition of $\Delta$ chosen in the literature varies (with values around 200); 
here unless otherwise specified we use the definition given by \cite{bryan_norman:98}, which characterizes the overdensity predicted for a virialized region that has undergone spherical collapse.

Within the radius of a dark matter halo there may be multiple, distinct peaks in the density field with virialized clumps of dark matter gravitationally bound to them. These {\it subhalos} are smaller than the {\it host} halo, and they orbit within the gravitational potential of the host halo. Resolving and tracking such objects is critical for making proper comparisons to the observed distribution of galaxies.

We note that the definition of halo radius given above, though common in the literature, may not be the most physically motivated definition of the boundary of a dark matter halo.  \cite{diemer_etal:13} have emphasized that the commonly used definitions of halo boundaries can lead to unphysical interpretations about halo mass accretion histories.  For example, measuring halo growth using $M_{\rm vir}$ will lead one to infer significant halo growth which is due just to the halo boundary being defined to larger radii with time, which they term ``pseudoevolution''.  Recently several authors have suggested an alternative concept, the ``splashback'' radius, which specifies the radius at which matter that is bound to the halo can orbit to after first collapse \citep{diemer_etal:14,more_etal:15,adhikari_etal:14,mansfield_etal:17}; this radius may also be more co-incident with the radius at which gas can shock heat, and at which infalling substructures can start being stripped by their host halos.  Because this has not been yet widely adopted in most of the studies we review, we do not adopt this convention here, but we note that it may change some of the detailed physical interpretation of results presented (it is not expected to change the qualitative conclusions).

\subsection{Empirical models of the galaxy--halo connection}
\label{sec:empirical}

The revolution in our understanding of the galaxy--halo connection has been driven by new physical insights as well as significant input from simple empirical models that connect observations from galaxy surveys to the predictions of the properties and evolution of dark matter halos in cosmological simulations. Here, we generally assume that the basic properties of dark matter halos are known for a given cosmological model.  They can be predicted directly using an N-body simulation, or using fitting functions that summarize the properties of halos in such a simulation.  To predict clustering statistics for example, one wants to know the abundance of dark matter halos (the ``halo mass function''; see e.g. \citealt{sheth_mo_tormen:01,tinker_etal:08_mf}), their clustering properties (the ``halo bias''; e.g. \citealt{sheth_tormen:99,tinker_etal:10_bias}, which can be a function of mass, redshift, and scale), the radial distribution of matter or substructures within halos, and the velocity distribution of dark matter or of substructures within halos.  In most of the discussion below, we assume these predictions are made with gravity-only N-body simulations; we explicitly discuss the impact of hydrodynamics and feedback where relevant.

\subsubsection{Abundance Matching}
Perhaps the simplest assumption one could make about the galaxy halo connection is that the most massive galaxies live in the most massive dark matter halos. This basic approach is generally called ``abundance matching'' in the literature (the most massive galaxy lives in the most massive halo; the second most massive galaxy lives in the next most massive halo, etc.)  The earliest versions of this assumption, applied before there were robust simulations that resolved cosmological structure {\em within} halos, assumed only one galaxy per halo \citep[e.g.][]{wechsler_etal:98,colin_etal:99,kravtsov_klypin:99,moustakas_somerville:02}.  However, CDM predicts structure on all scales, and thus predicts that dark matter halos host distinct substructures.  These substructures (above a certain mass) are expected to host galaxies.  A simple {\em ansatz} is thus that each halo {\em and subhalo} hosts a galaxy, with the mass or luminosity of a galaxy matched by abundance to the mass or velocity of the dark matter (sub)halo in which it lives; this is often referred to as subhalo abundance matching (``AM'' or ``SHAM'') in the literature \citep{kravtsov_etal:04, tasitsiomi_etal:04,vale_ostriker:04}.

Once this assumption is made, one can calculate a range of statistics for the model.  Although the earliest versions of these models were sometimes referred to as zero parameter models, there are in fact a set of assumptions or parameters that need to be specified.  The two most important are: (1) what halo property is best matched to what galaxy property? and (2) what is the scatter between these properties?  It was realized quickly that while subhalos are rapidly stripped of their outer material after being accreted into a larger dark matter halo, galaxy stripping starts much later \citep{nagai_kravtsov:05}.  Thus, one might expect a model which matches galaxies to halo properties at the time they are accreted into their host halos to provide a better match to a luminosity-selected galaxy sample; this was demonstrated by\cite{conroy_etal:06}.  Later work has investigated several alternative possibilities for the matching proxy, discussed further in \S \ref{sec:bestmatchhalos}.  However, even in the presence of scatter between galaxy and halo masses, abundance matching is best thought of as a non-parametric technique that directly connects the stellar mass function to the halo mass function \citep{tasitsiomi_etal:04}. This can be done by deconvolving the scatter, as described by \cite{behroozi_etal:10}; see in particular \S 3.3.1 of that work for the equations governing this deconvolution .\footnote{A code to implement this procedure is available at {\tt http://bitbucket.org/yymao/abundancematching}}  The consequences of and constraints on scatter are discussed further in \S \ref{sec:scatter}.  We note that modern versions of abundance matching models generally require high-resolution simulations; they depend on resolved substructure and on accurate merger trees to track the path of halos at least to the point in time that they started being tidally stripped.

\subsubsection{The stellar mass/halo mass relation (SHMR)}
\begin{figure}[t]
\includegraphics[width=0.9\textwidth]{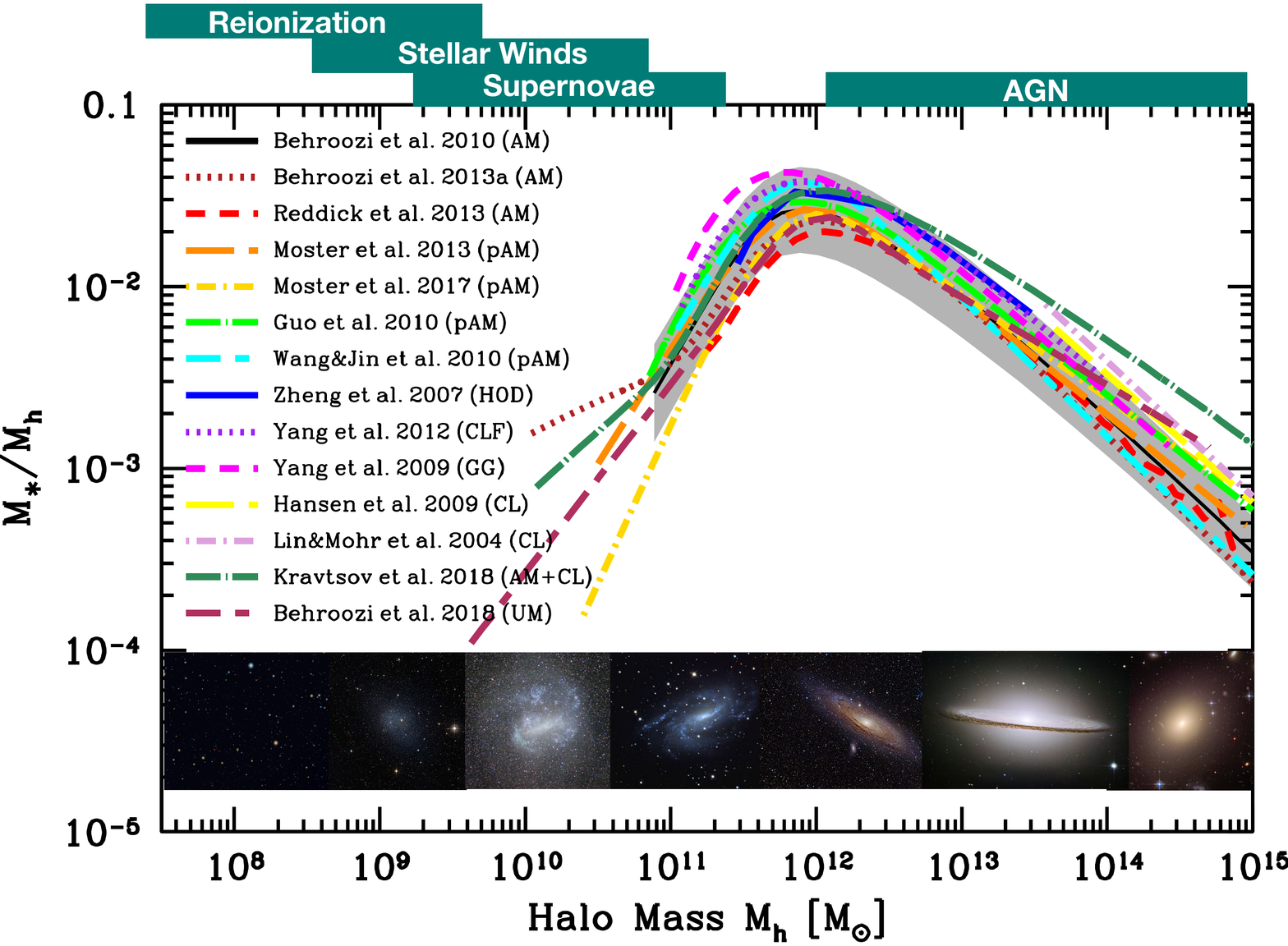}
\caption{\footnotesize{The galaxy stellar mass-to-halo mass ratio of central galaxies at $z=0$.  The figure (based on data compiled in
\citealt{behroozi_etal:17}) shows constraints from a number of different methods: direct abundance matching \citep{behroozi_etal:10, reddick_etal:13, behroozi_etal:13}; ``parameterized abundance matching,'' in which this relationship is parameterized and then those parameters are fit with the stellar mass function and possibly other observables \citep{guo_etal:10,wang_jing:10,moster_etal:10, moster_etal:13}; 
from modeling the halo occupation distribution \citep{zheng_etal:07} or the CLF \citep{yang_etal:09} and constraining it with two-point clustering; by direct measurement of the central galaxies in galaxy groups and clusters \citep{lin_mohr:04, yang_etal:09, hansen_etal:09, kravtsov_etal:14}; and the ``Universe Machine,'' an empirical model that traces galaxies through their histories \citep{behroozi_etal:17}.
Bottom panel shows example galaxies that are hosted by halos in the specified mass range. On the top of the figure, we indicate key physical processes that may be responsible for ejecting or heating gas or suppressing star formation at those mass scales. {\em Figure adapted from \cite{behroozi_etal:17} with permission.}}}
\label{pretty_gmhmr}
\end{figure}

Abundance matching can be used to determine the typical galaxy stellar mass at a given halo mass, or galaxy stellar-to-halo mass relation, which we abbreviate as SHMR.  An alternative to inferring this SHMR from non-parametric abundance matching is to parameterize it and constrain the parameters \citep[e.g.][]{moster_etal:10}.  The SHMR for central galaxies is shown in \textbf{Figure \ref{pretty_gmhmr}}, as constrained by non-parametric abundance matching, as inferred by a parametric SHMR constrained by abundance and clustering data, and as derived by a number of other methods that will be described below.   The basic shape of this relation derives from the mismatch between the halo mass function and the galaxy stellar mass function or luminosity function, which declines rapidly below typical galaxies and has a much shallower faint-end slope than the halo mass function.  One can see several clear features in this relation, which are identified consistently using any of the methods used to constrain it.  First, the peak efficiency of galaxy formation is always quite low: if all halos are assumed to host the universal baryon fraction $\Omega_b/\Omega_m$ of $17\%$, at its maximum, these results show that just $\sim 20$--$30$\% of baryons have turned into stars, resulting in a SHMR that peaks at just a few percent.  This maximum galaxy formation efficiency occurs around the mass of halos hosting typical $L*$ galaxies like the Milky Way, around $10^{12} \msun$; we refer to this as the pivot mass.  At higher and lower masses, galaxy formation is even less efficient.  Roughly, the stellar mass of central scales as $\mgal \sim \mhalo^{2-3}$ at dwarf masses and $\mgal \sim \mhalo^{1/3}$ at the high mass end.  Images of typical galaxies that populate halos of a given mass are shown below the relation.  

\begin{marginnote}[]
\entry{SHMR}{The stellar-to-halo mass relation.  This can be predicted with models of galaxy formation, inferred from parameterized models, or measured directly.}
\end{marginnote}

This decrease in the efficiency of star formation is a signature of strong feedback processes from the formation of stars and black holes. It is likely due to combination of a number of processes:  at high mass, AGN feedback can act to heat halo gas and limit future star formation \citep{silk_rees:98,croton_etal:06}; 
at low mass, feedback from massive stars is believed to be important in driving winds that eject gas, or prevent it from coming into a galaxy \citep{dekel_silk:86,hopkins_etal:12}; at even lower masses, galaxies can be too small to hold onto their gas during the reionization period around $z \sim 6$ \citep{bullock_etal:00}.  We discuss the constraints on this relation in more detail in Sections \ref{sec:constraints} and \ref{sec:constraints_secondary}, but here we note that many different techniques are telling the same basic story.  

Below some threshold halo mass, galaxies will no longer be able to form at all.  The smallest known galaxies, ultra-faint dwarf galaxies, have measured dynamical masses in their inner regions larger than a few times $10^7$ $\msol$, which is most likely equivalent to halo virial masses of larger than $10^9$ $\msol$.  The exact value of the minimum mass at which a halo can host a galaxy is still somewhat uncertain, as is the slope of and scatter in SHMR for halos below $\sim 10^{11}$ $\msol$.  Each of these has important consequences for understanding the lowest mass galaxies, and also has implications for the nature of dark matter \citep{bullock_boylankolchin:17}.

We note that the SHMR generally parameterizes $\mgal$ as a function of $\mhalo$. Due to scatter in these two quantities, quantifying the galaxy--halo connection with the mean halo mass in bins of $\mgal$--- as done observationally --- does not yield the same mean relation. We discuss this in detail in \S \ref{sec:scatter}.

\subsubsection{The Halo Occupation Distribution and Conditional Luminosity Function}
A popular way to describe the relationship between galaxies and dark matter halos is through the Halo Occupation Distribution (HOD), which specifies the probability distribution for the number of galaxies meeting some criteria (for example, a luminosity or stellar mass threshold) in a halo, generally conditioned on its mass, $P(N|M)$.  Typically this PDF is quantified separately for the central galaxies of halos and the satellite galaxies that orbit within the halos. For the former, a Bernoulli distribution is assumed, while for satellites a Poisson distribution is assumed. Under these assumptions the standard HOD is thus fully characterized by its mean occupation number $\langle N|M\rangle$; we discuss this assumption in Section \ref{sec:poisson}. In principle, the HOD can be a function of properties other than halo mass; we discuss this possibility in \S \ref{sec:complications}.  

The connection between modern halo occupation models and measurements of galaxy clustering started to be explored by several workers in the early 2000s \citep[e.g.][]{Peacock2000,Seljak2000,benson_etal:00,wechsler_etal:01,scoccimarro_etal:01,berlind_weinberg:02,bws:02}, and is now well constrained for a wide range of galaxy samples. The functional form of the HOD for mass- or luminosity-selected galaxies is generally assumed to be similar to that of dark matter subhalos within their hosts. This was first studied in detail by \cite{kravtsov_etal:04}, who found that the HOD for samples of subhalos is well-described by a power law of subhalos $N \sim M$, with the addition of a central galaxy; for a given threshold on galaxy stellar mass, a typical central galaxy can be found in halos 10--30 times less massive than the halos that host satellite galaxies of the same stellar mass (examples are shown in the next section). This rough functional form has been shown to hold for luminosity-threshold or stellar mass-threshold samples of galaxies. In general, such an HOD can be described by 3--5 parameters for a given galaxy sample.  Commonly used parameterizations are given in \cite{zheng_etal:05} (their equations [1] and [3]) and \cite{reddick_etal:13} (their equations 9 and 10). For more complicated galaxy samples (e.g. selected by star formation rates, colors, or emission lines), the functional form of the HOD can be significantly more complicated (e.g., \citealt{skibba_sheth:09}).

The conditional luminosity function (CLF) and conditional stellar mass function (CSMF) go one step further to describe the full distribution of galaxy luminosities for a given halo mass.
It is generally described separately by the distribution of central galaxy luminosities $P(L_c|M)$ and satellite galaxy luminosity functions $\Phi(L_{\rm sat}|M)$. This can be inferred directly from measurements of groups and clusters \citep{lin_etal:04,weinmann_etal:06, yang_etal:08,hansen_etal:09, yang_etal:09} or from a full model for galaxy clustering and abundance \citep{yang_etal:03, cooray:06}.  In general, this parameterization distinguishes between central galaxies, which are usually assumed to follow a lognormal distribution of stellar masses or luminosities at fixed halo mass, and satellite galaxies, which are usually assumed to follow a Schechter function \citep{schechter:76} whose parameters scale with halo mass. A concise review of the equations governing the CLF can be found in \S 3.7 of \cite{vandenbosch_etal:13}.

For both the CLF and the HOD, model predictions can be made in two ways. Both models specify the number of galaxies per halo, thus one can populate halos identified in an N-body simulation using a Monte Carlo approach, and `measure' observables from the resulting mock galaxy catalog. Alternatively, both of these frameworks can be combined with an analytic halo model of dark matter clustering to make predictions for some statistics analytically (see, e.g., \citealt{tinker_etal:05} and \citealt{vandenbosch_etal:13}). The CLF and HOD parameterize the galaxy--halo connection differently, but in spirit they quantify the same thing. Either method can be used to quantify the other (see, e.g., \citealt{leauthaud_etal:11a}).

\begin{marginnote}[]
\entry{HOD}{Halo occupation distribution. This specifies the probability distribution for the number of galaxies in a halo, generally conditioned on its mass, $P(N|M)$.}
\entry{CLF}{Conditional luminosity function. This specifies the luminosity function of galaxies (both centrals and satellites) conditioned on halo mass.  
}
\end{marginnote}

\subsubsection{Empirical Modeling of Galaxy Formation Histories}

Somewhat intermediate to the abundance matching and HOD/CLF models that describe a galaxy population at a fixed epoch and the full semi-analytic approach described in \S \ref{sec:sams} is a class of models that trace galaxies within their dark matter halos over time, but directly constrain the galaxy--halo connection at each epoch.  \cite{conroy_wechsler:09} developed a simple approach along these lines using abundance matching at each epoch to determine the SHMR, combined with the typical mass accretion histories to connect halos through time, to determine typical galaxy accretion histories and star formation histories across cosmic time.  \cite{behroozi_etal:13}, and \cite{moster_etal:13} extended this work using simulated mass accretion histories (following on earlier work from \citealt{yang_etal:12} with analytic approximations for halo properties) as well as updated constraints from the evolution of the galaxy stellar mass function and galaxy star formation rates to put strong constraints on the typical trajectories of galaxies through time.

This approach is being taken further by many workers \citep{becker:15, rodriguez-puebla_etal:16, cohn:17, moster_etal:17, behroozi_etal:17}; instead of parameterizing the connection between galaxy stellar mass and halo properties at a given epoch, one can parameterize for example the relationship between the galaxy star formation rate and the halo mass accretion rate, and then trace these histories through time using simulated merger histories.  This is a powerful approach which allows one in principle to use a range of data to constrain the model, and to make predictions for the distribution of galaxy star formation histories as well as their statistical properties at any epoch.  In general this approach also requires high-resolution simulations to construct robust merger trees of dark matter halos and to trace the evolution of subhalos. 

\subsection{Physical models of galaxy formation}
\label{sec:physical}
Physical models of galaxy formation attempt to either directly simulate or to model the basic physical processes in galaxy formation.  The current status and approaches of these models, including both hydrodynamical simulations and semi-analytic models, were recently reviewed by \citet{somerville_dave:15}.  Here we primarily focus on the connection to and contrast with empirical models, as well as the interplay between these various approaches.

\subsubsection{Hydrodynamical Simulations}
Hydrodynamical simulations model galaxy formation by solving the equations of gravity and hydrodynamics in a cosmological context, incorporating such processes as gas cooling, stellar-feedback driven winds, and feedback from black holes and supernovae, and in some cases magnetic fields and cosmic rays, and tracing the properties of dark matter, gas, and stars in given resolution elements over time. Although they contain extensive physical prescriptions, they cannot simulate the full range of scales needed for galaxy formation in a cosmological context without some parameterizations below the resolution scale, generally termed ``subgrid physics.''  These subgrid physics parameterizations need to be tuned, either through direct tests with observations or by comparison to constraints with empirical models that connect observations to dark matter halos.

Although there is still significant uncertainty in the details, there has been dramatic progress in producing realistic galaxy populations in hydrodynamical simulations over the past decade, due to increasing resolution as well as improved physical models for star formation and feedback, based on insight from a wide range of observations.  These models provide our best understanding of the complex interplay between the physical processes of galaxy formation, and they can thus be used to inform and test the assumptions of empirical models.  The earliest studies of the halo occupation in hydrodynamical simulations were performed before it was well-constrained by empirical models \citep{White2001,Pearce2001,berlind_etal:03}, following earlier work looking at the occupation in a semi-analytic model by \cite{benson_etal:00}.
These studies provided useful insight into early modeling approaches for the HOD, e.g. \cite{zheng_etal:05} used smoothed particle hydrodynamics simulations and semi-analytic models to propose forms for the HOD and CLF that were later constrained with the best-available clustering data from the Sloan Digital Sky Survey.  More recently, \cite{simha_etal:12} and \cite{chaves-montero_etal:16} have tested the key assumptions of the subhalo abundance matching approach with modern cosmological hydrodynamical simulations.

The interplay goes both ways: in recent years, measuring the galaxy--halo connection either through the SHMR or the halo occupation in these simulations and comparing to constraints obtained from empirical models and/or combinations of data has become a standard test for cosmological hydrodynamical simulations and semi-analytic models \citep{genel_etal:14, vogelsberger_etal:14, schaye_etal:15}.
Because these models are computationally expensive (generally, at least an order of magnitude more CPU time to simulate a given volume than dark matter only simulations), the SHMR and other parameterizations of the galaxy--halo connection provide very useful intermediate targets that can be easier to match than full forward modeling of the entire galaxy population and comparing directly to the range of observables that have been used to constrain it. This can be done for full cosmological simulations \cite[e.g.][]{crain_etal:15}, or one can even run a small set of high-resolution resimulations and evaluate whether the typical galaxy mass agrees with that inferred from empirical models \citep{stinson_etal:13, munshi_etal:13}.
Below, we review comparisons between these models and current empirical constraints.

\subsubsection{Semi-analytic models of galaxy formation}
\label{sec:sams}

Semi-analytic models of galaxy formation \citep{white_frenk:91, kauffmann_etal:93, somerville:99, cole_etal:00, bower_etal:06, guo_etal:13} 
aim to model the same basic processes of galaxy formation in a computationally efficient manner, by approximating the various physical processes with analytic prescriptions that are traced through the merging history of dark matter halos.  In current models, these prescriptions are most often traced through merger trees extracted from N-body simulations.  Although these models are significantly less computationally expensive than hydrodynamical simulations, they generally have a large number (10--30) of parameters and fully exploring this parameter space has remained a challenge.  These prescriptions also necessarily make simplifying assumptions, that need to be continually tested both with full hydrodynamical simulations and with data.  Several recent studies have used Monte Carlo Markov chain techniques to directly constrain the semi-analytic model parameter space with data \citep{henriques_etal:09, lu_etal:11, lu_etal:14, henriques_etal:15}.  Fully constraining these models with clustering data and other spatial statisics is still challenging due to the large parameter space and computational expense.   As an alternative, the SHMR and other aspects of the parameterized galaxy--halo connection can provide useful intermediate steps to test the agreement of models with a wide range of data.

\begin{marginnote}[]
\entry{SAM}{Semi-analytic model}
\end{marginnote}

\subsection{Complementarity between approaches}
One of the most encouraging aspects of the current state of galaxy formation modeling is that each of the approaches outlined above is increasingly being used to inform the others: more direct physical models can be used to inform and test the parameterizations and assumptions of empirical approaches, and empirical constraints can be used to efficiently synthesize diverse constraints from data and pin down uncertainties in the physical parameterizations.  At present, due to the computational expense of more physical models, empirical models also are more widely used in studies that jointly constrain the galaxy--halo connection with cosmological parameters.  Empirical models are also important for cases in which one wants to marginalize over possible uncertainty in the galaxy--halo connection in order to robustly infer cosmological parameters or uncertain dark matter physics.
\section{MEASUREMENTS THAT INFORM  THE GALAXY--HALO CONNECTION}
\label{sec:observations}

In the previous section we discussed both empirical and physical models for the galaxy--halo connection, and the interplay between them.  Here, we review the most important measurements that are currently used to inform the galaxy--halo connection: galaxy abundances (\S \ref{sec:abundance}), galaxy clustering (\S \ref{sec:clustering}), group and cluster catalogs (\S \ref{sec:groups}), weak gravitational lensing (\S \ref{sec:lensing}), and additional observables including spatial statistics and scaling relations (\S \ref{sec:additional}). For a given cosmological model and galaxy formation model, the abundance of objects, the relationship between satellites and centrals, and the spatial distribution of galaxies are related through their galaxy--halo connection.  This is true for any galaxy--halo connection, whether it derives from physical models or a parameterized functional form.  In general, modeling approaches that make more physical assumptions (towards the left in {\bf Figure \ref{models}}) are more predictive than more data-driven approaches.  However, some empirical models make specific assumptions such that when we use them to describe one or a small set of observables, they immediately make predictions for a large set of other observables.

\begin{figure}[h]
\includegraphics[width=0.9\textwidth]{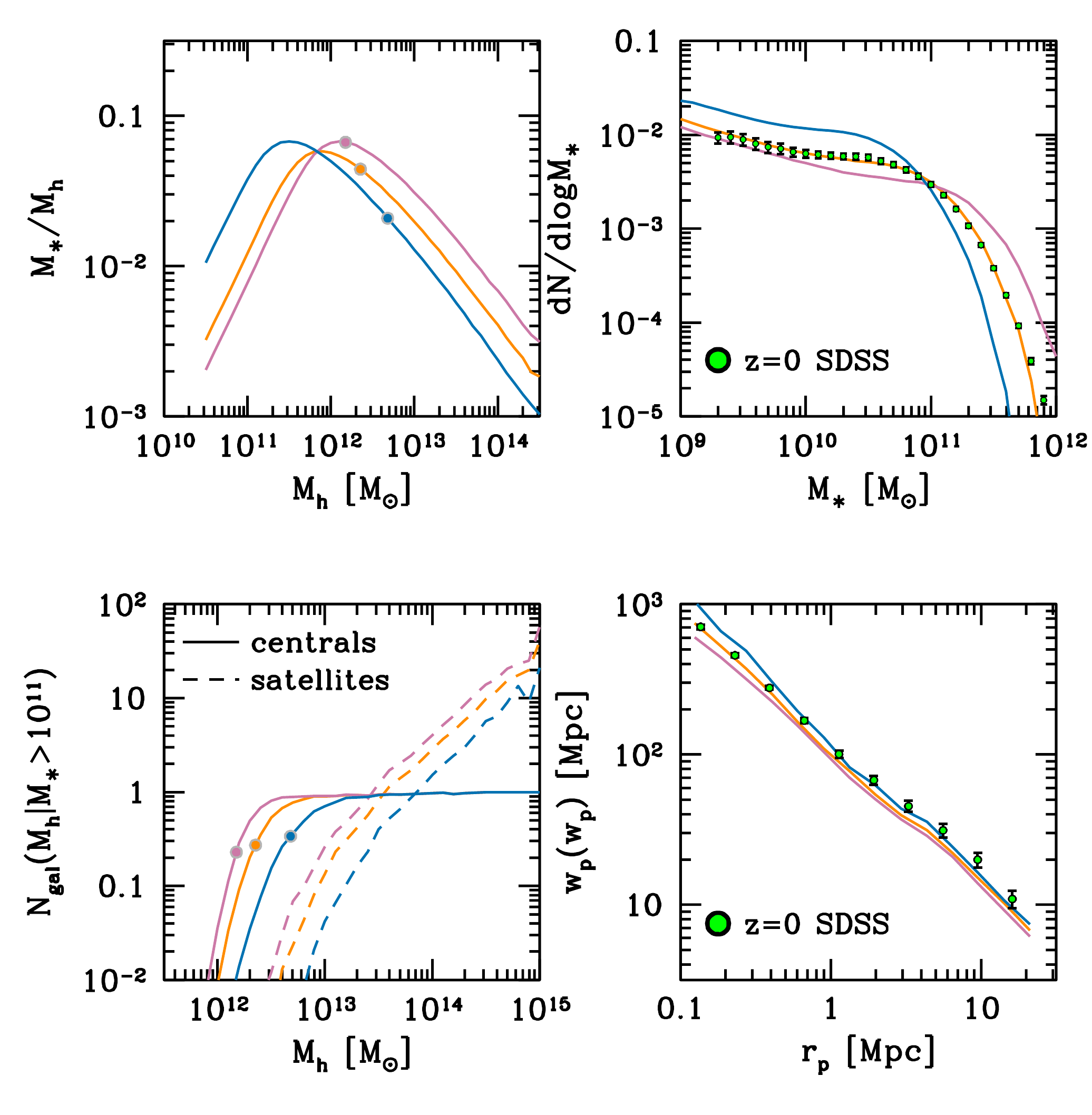}
\caption{Examples of the SHMR. {\em Upper left:} The $\mgal/\mhalo$ ratio for three parameterized SHMRs. The middle {\em (orange)} curve is constrained to match the stellar mass function at $z=0$ ({\em Upper Right}), whereas the other models {\em(blue and green curves)} show the sensitivity of the mass function to changes in the SHMR. The filled circles in each curve indicate the halo mass at which the mean $\mgal$ is $10^{11}$ $\msol$. {\em Bottom left:} The mean halo occupation function for galaxies with $\mgal>10^{11}$ $\msol$ predicted by each SHMR model. Here, the filled circles show the same halo mass as the circles in the upper left. {\em Bottom right:} Predictions for the projected correlation function of these galaxies, with SDSS measurements for comparison. All model calculations are performed using the public halo catalogs from the Bolshoi-Planck simulation (\citealt{klypin_etal:16}).
Abbreviations: SDSS, Sloan Digital Sky Suryve; SHMR, stellar-to-halo mass relation.
}
\label{pedagogical1}
\end{figure}

\subsection{Galaxy abundance}
\label{sec:abundance}

Given the assumption that galaxy properties and halo properties are closely connected, the most important constraint on the galaxy--halo connection for a given cosmological model (which in turn predicts the abundance of dark matter halos) is the abundance of galaxies as a function of stellar mass or luminosity, i.e., the stellar mass function (SMF) or the luminosity function.  The SMF inferred from the measurements of galaxy abundance in the local universe is now a very statistically precise measurement\citep{bell_etal:03,li_white:09,baldry_etal:12, bernardi_etal:17}, but there are still a number of important systematic issues that impact its normalization and mass scaling and can have consequences for the galaxy--halo connection and its interpretation.  Constraints on the evolution of galaxy stellar mass functions have also improved significantly over the past decade \citep{perez_etal:05,moustakas_etal:13}, leading to consequential improvements in the evolution of the galaxy--halo connection.  

As discussed in \S 2, abundance matching can be done directly, in a non-parametric way that reproduces the observed galaxy SMF (or the luminosity function) by construction. Alternatively, it can be done parametrically, with a function that maps halo mass---or alternative halo properties---onto galaxy stellar mass. The free parameters are then constrained by the measurements of the SMF.

The relationship between galaxy abundance, galaxy clustering, the halo occupation, and the galaxy-to-halo mass ratio for this parameterized model is shown in \textbf{Figure \ref{pedagogical1}}. 
The parameterization used here is taken from \cite{behroozi_etal:10}, with the main features being characteristic mass scales for $\mgal$ and $\mhalo$, which constitute the location of the pivot point in the SHMR, and power-law slopes above and below the pivot point.  A model that describes the abundance data well is shown in {\bf Figure 3b}. These data were measured independently for the present work. They represent the SMF at $z=0$ based on data from SDSS DR7 (\citealt{dr7}) using the stellar masses of \cite{chen_etal:12}. Other curves show models that have pushed the pivot point too high and too low. These models clearly fail to reproduce the abundance of galaxies: moving the pivot point too low yields too many low-mass galaxies and fails to produce any at the high mass end, while a pivot point too high has the opposite effect. The filled circles on each SHMR curve show the halo mass scale at which the galaxy mass is $M_* = 10^{11}$ $\msol$, on average. In all of these models, we have set the log-normal scatter in $\log\mgal$ at fixed $\mhalo$ to be 0.2 dex. 
Modifying the power-law slopes of the SHMR will have a similar effect: altering the SMF slope at the high or low mass end will change the slope of the SHMR.  The general shape and amplitude of the fit produced in \textbf{Figure \ref{pedagogical1}} echoes the compilation of results shown in \textbf{Figure \ref{pretty_gmhmr}}---a pivot point at $\mhalo\sim 10^{12}$ $\msol$, a steep slope at lower masses and a shallower slope at higher masses.  We note that at low masses (stellar masses below $\sim 10^{9} \msun$), there is evidence of a break in the SMF power law to produce more galaxies than a fixed slope would produce, leading to an increasing SHMR at the low mass end.

\subsection{Two-point galaxy clustering}
\label{sec:clustering}

Because the abundance of dark matter halos is connected to their clustering properties, a given abundance matching or parameterized SHMR model makes a prediction for the clustering of galaxies. Once the parameters of the model have been set---the scatter, the abundance matching variable---there is no more freedom in the model after fitting for the galaxy abundance. The bottom panels of \textbf{Figure \ref{pedagogical1}} show the predictions each of the SHMR models make for the mean occupation function and clustering of galaxies with $\mgal\ge 10^{11}$ $\msol$. The filled circles in the bottom left correspond to the same $\mhalo$ as the filled circle in the top left panel. Although these are non-parametric estimates of the HOD, the shapes are well-matched by the functional forms for central and satellite galaxies discussed earlier. Increasing or decreasing the halo mass scale at $\mgal=10^{11}$ $\msol$ increases and decreases the overall mass scale of the HOD. This has the impact of increasing and decreasing the clustering of galaxies. 

For HOD and CLF models, abundances alone cannot constrain the parameters of the models. This is because the number of satellites within a halo is a parameterized function, rather than mapped onto the substructure in a simulation. Thus, measurements that provide information about the halo mass are necessary to constrain the parameters of the HOD and CLF. Galaxy clustering is perhaps the most commonly used; galaxy clustering at $r<1$ Mpc is highly sensitive to the fraction of galaxies that are satellites because the number of pairs within a halo increases with the square of the number of satellites. Clustering at larger scales is most sensitive to the overall halo mass scale and the scatter between halo mass and the observable quantity by which galaxies are selected. The halo occupation framework predicts a transition scale in the clustering of galaxies at $r\sim 1$ Mpc. This transition scale naturally explains the deviations from a power-law measured in the clustering of galaxies \citep{zehavi_etal:04,zehavi_etal:05}, as well as its evolution as a function of galaxy luminosity and redshift \citep{conroy_etal:06}. A convenient way to quantify large-scale clustering of a sample of objects is by their bias relative to dark matter. Bias generically is related to the ratio between the clustering of the objects and that of the dark matter: $b^2 = \xi/\xi_m$, where $\xi$ is the two-point correlation function of galaxies and the subscript $m$ denotes matter. Bias is usually quantified at large scales, i.e., $r>10$ Mpc. The bias of dark matter halos has a non-linear dependence on $\mhalo$---low mass halos have a roughly constant bias, while at higher masses bias rises rapidly with increasing $\mhalo$. Thus, for the three different models presented in \textbf{Figure \ref{pedagogical1}}, the bias of the model increases monotonically with the characteristic halo mass indicated by the filled circles.

This does not mean that clustering has no information to offer for abundance-matching models. As we discuss in \S 4, models with different assumptions regarding how galaxies are matched to halos can change the relative rank-ordering of halos and subhalos, thus shifting the relative populations of central and satellite galaxies.  Scatter can also not be constrained by abundances alone.  The only way to discriminate between these models is to bring in information other than galaxy abundances, from clustering or one or more of the other observables described below (see further discussion in \S \ref{sec:scatter}).

\subsection{Group and cluster catalogs}
\label{sec:groups}

An alternative method to constrain the galaxy--halo connection is  to find individual dark matter halos observationally and measure the galaxy content within them. Galaxy clusters are relatively easy to find in observational data owing to their large mass and the large number of galaxies contained within them. \textbf{Figure \ref{pretty_gmhmr}} includes measurements of the SHMR at the cluster mass scale from \cite{kravtsov_etal:14}
using X-ray observations to detect clusters and \cite{hansen_etal:05} using optical imaging data. Optical cluster catalogs can now also be extended to very large volumes and to significantly higher redshift, for example using red sequence techniques like RedMaPPer \citep{rykoff_etal:14, rykoff_etal:16}; given the statistical power of these data sets, the primary challenge in using these to infer the galaxy--halo connection is a full modeling of systematics in the mass--observable relation.

Modern day galaxy group catalogs are able to probe much lower halo mass scales. The group-finding algorithm of \cite{yang_etal:05} uses a variation of the abundance matching ansatz to assign dark matter halo masses statistically to all galaxies within a given sample. After an initial guess about which galaxies belong to groups and which do not, abundance matching is performed on the {\it total} group stellar mass and the host dark matter halo mass function. Each galaxy is assigned a probability of being within a given dark matter halo, using these updated halo masses, and the process is iterated until convergence. The resulting halo occupation statistics that are derived from the final group catalog are in general agreement with those inferred from an HOD analysis of clustering (\citealt{yang_etal:08,tinker_etal:11,campbell_etal:15}). The application of this algorithm to SDSS data in \cite{yang_etal:09} is also presented in \textbf{Figure \ref{pretty_gmhmr}}. Groups can also be used in combination with two-point clustering (\citealt{sinha_etal:17}).

\subsection{Weak gravitational lensing}
\label{sec:lensing}

It is straightforward to extend implementations of the galaxy--halo connection from models of the two-point clustering of galaxies to the cross-correlation between galaxies and matter \citep[e.g.][]{tasitsiomi_etal:04,yoo_etal:06, cacciato_etal:09, leauthaud_etal:12}. Generally, the observational quantity being modeled is $\DS$, the excess surface mass density at a projected distance $R_p$ away from a galaxy, which is related to the galaxy-mass correlation function $\xi_{gm}$. If all galaxies were central, then $\DS$ would simply be a measure of the projected dark matter halo profiles of all the galaxies in the sample being modeled. Satellite galaxies complicate the interpretation of the lensing signal, requiring a full model of the galaxy--halo connection. 

Lensing generally has lower signal-to-noise than two-point clustering, but the constraints offered by a lensing analysis are complementary to that of an analysis only using clustering. Lensing directly measures the mass of the dark matter halos around galaxies, whereas in a clustering analysis the masses are inferred indirectly from the relationship between halo mass and clustering, as well as the constraints on the abundance of halos in a given cosmological model. As we discuss in \S \ref{sec:assbias}, properties other than halo mass may influence clustering. Thus, a cross comparison of constraints on halo occupation from clustering and lensing is a necessary check on possible systematic errors or secondary parameters compared to using clustering alone. 

\subsection{Additional observables}
\label{sec:additional}

Although the abundance of galaxies and their two-point clustering are the methods that have been most commonly used to statistically constrain the galaxy--halo connection, there is information on this relationship from nearly any measure of the spatial distribution of galaxies. Within the literature, these include satellite kinematics (\citealt{vandenbosch_etal:04,more_etal:11}), galaxy voids (\citealt{sheth_weygaert:04,furlanetto_piran:06,tinker_etal:08_voids}), counts-in-cells (\citealt{benson_etal:00,berrier_etal:11}), and three-point statistics \citep{marin_etal:08, guo_etal:15}.  Measurements of the intracluster light around clusters and groups of galaxies can also provide important constraints that distinguish the merging and growth history of galaxies \citep{conroy_etal:07}.

Scaling relations of the internal properties of galaxies, such as Tully-Fisher and Faber-Jackson \citep{tully_fisher:77,faber_jackson:76}, also offer information to the extent that dark matter influences the dynamics of visible matter \citep{dutton_etal:10,cappellari_etal:13,desmond_wechsler:15,desmond_wechsler:17}.  At low masses, where large samples with observable spatial statistics are not yet available, these scaling relations provide the primary information about the galaxy--halo connection \citep{mcconnachie_etal}.  Strong lensing can provide additional information about the masses of elliptical galaxies, at least in their inner regions \citep{sonnenfeld_etal:15}.

\section{BEYOND THE SIMPLEST MODELS OF THE GALAXY--HALO CONNECTION}
\label{sec:complications}

The simplest versions of the empirical models we outlined in \S \ref{sec:models}, which connect one primary galaxy property to one halo property, describe the basic relations very well.  However, they have now been shown in the literature to be inadequate for describing both the spatial distribution of galaxies at the precision it can now be measured and the full richness of galaxy properties and correlations that are observable. Here, we discuss additional modeling aspects that may be required. We first discuss the question of which halo property should be primary (\S \ref{sec:bestmatchhalos}).  We then consider the question of alternative primary galaxy properties (\S \ref{sec:bestmatchgals}).  We then review other aspects of the galaxy--halo connection that can impact clustering properties, even for a given SHMR: scatter between galaxy mass and halo mass (\S \ref{sec:scatter}), assembly bias (\S \ref{sec:assbias}), and the ratio of galaxy mass to halo mass between centrals and satellites (or, equivalently, the satellite fraction as a function of stellar mass; \S \ref{sec:complications_censat}). We then discuss properties of the satellite distribution (\S \ref{sec:poisson}), and consider approaches to joint modeling of additional galaxy properties, including galaxy star formation rates, galaxy sizes, and gas properties.

\subsection{What halo property is best matched to galaxies?}
\label{sec:bestmatchhalos}
The existence of a galaxy--halo connection does not specify which {\it galaxy} property correlates with {\it which} halo property, though mass is a typical assumption for both. Mapping galaxy stellar mass to different halo properties will yield the same SMF, but different clustering signals. \cite{kravtsov_etal:04} originally proposed using the maximum circular velocity of halos $\vmax$ ---both parent halos and subhalos---to match onto galaxies. Whereas the mass of a subhalo is subject to intense tidal stripping immediately upon entering a larger halo, $\vmax$ is more robust to stripping. A solution to the tidal stripping problem that allows one to use $\mhalo$ is to use $\mhalo$ at the time the subhalo was accreted in order to match to galaxy properties. These ideas can be combined in various ways, all of which yield slightly different results in terms of the spatial distribution of galaxies.

\begin{figure}[h]
\includegraphics[width=5in]{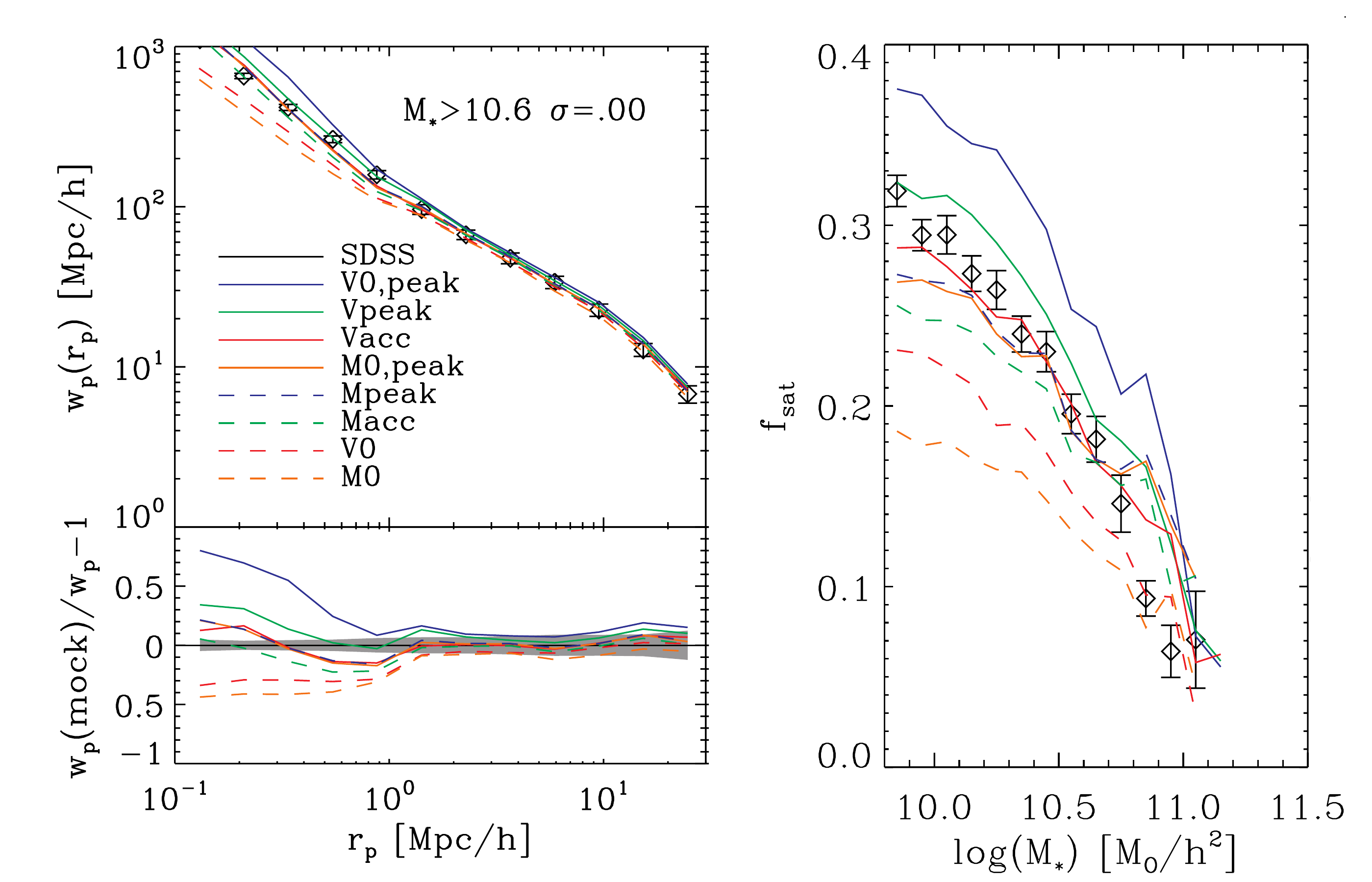}
\caption{\footnotesize{{\it Left Panel:} Projected galaxy clustering as a function of scale, for different variants of abundance matching models, where the galaxy properties are matched to different halo properties. In these models no scatter is assumed. Points with error bars are measurements from SDSS. {\it Right Panel:} The fraction of galaxies that are satellites as a function of $\mgal$. The line types are the same as in the left panel. Points with error bars are from the SDSS galaxy group catalog of \cite{tinker_etal:11}. All theoretical models have been passed through the group finder to account for any biases in this process. Both panels are from \citet{reddick_etal:13}.}}
\label{reddick13}
\end{figure}

\begin{textbox}[h]\section{Different Halo Properties for Abundance Matching}
\begin{itemize}
\item {\bf $M_0$:} The halo mass today or at the epoch being modeled.
\item {\bf $M_{\rm acc}$:} Mass at the time of accretion. Equivalent to $M_0$ for host halos.
\item {\bf $M_{\rm peak}$}: In the entire history of a halo, the highest mass achieved.
\item {\bf $V_0$}. Maximum circular velocity of the halo today or at the epoch being modeled.
\item {\bf $V_{\rm acc}$:} The $\vmax$ at the time of accretion. Equivalent to $\vmax$ for host halos.
\item{\bf $V_{\rm peak}$:} In the entire history of a halo, the highest  $\vmax$ achieved.
\item {\bf $M_{0,\rm peak}$ and $V_{0,\rm peak}$:} For subhalos, $M_{\rm peak}$ or $V_{\rm peak}$. For host halos, these values are taken at the epoch being modeled.
\item{{\bf $V^\alpha:$} a smooth transition between {\bf $M_{\rm peak}$} and  {\bf $V_{\rm peak}$:} $v_{\text{vir}}\left(\frac{v_{\text{max}}}{v_{\text{vir}}}\right)^\alpha$, with $0 \le \alpha \le 1$
}
\end{itemize}
\end{textbox}

Each of the halo properties in the sidebar titled Different Halo Properties for Abundance Matching can be used, either in a non-parametric abundance matching model or in a parameterized SHMR, to map galaxies onto halos to match the observed luminosity function or SMF. Each of these models will produce slightly different spatial distributions, quantified by the two-point correlation functions shown in \textbf{Figure \ref{reddick13}}.  In general, the models that use $\vmax$, in some form or another, produce galaxy samples that are more highly clustered than models that use $\mhalo$. This is due to two effects which can both impact the clustering: assembly bias, and the different relationships between the properties of centrals and satellites. We discuss the former in \S \ref{sec:assbias}. For the latter, the impact of different abundance matching quantities is clearly seen in the right-hand side of \textbf{Figure \ref{reddick13}}, which shows the satellite fraction, $\fsat$, for each of the models as a function of galaxy stellar mass. Models with a higher $\fsat$ yield a higher clustering amplitude, especially at smaller scales ($r_p< 1$ Mpc).

\subsection{What galaxy property is best matched to halos?}
\label{sec:bestmatchgals}
Through most of this review, we consider galaxy stellar mass to be the primary galaxy property that is most tightly correlated with the properties of the galaxy halos.  However, it is interesting to consider other possibilities. In the literature, galaxy stellar mass and galaxy luminosity have been treated rather interchangeably.  It is likely that observational uncertainties in measuring galaxy masses and luminosities may be larger than the difference in the scatter between them, but with higher precision measurements it may be possible to make this distinction.  The total stellar mass or luminosity in groups and clusters of galaxies is better correlated with halo mass than is the stellar mass or luminosity of the central galaxy; the former scales roughly as $M_{\rm *,tot} \sim M_{h}^{2/3}$ on the massive end, whereas the latter scales roughly as $M_{\rm *,cen} \sim M_{h}^{1/3}$ on the massive end \citep[e.g.][]{lin_mohr:04}.

Most of this review assumes that stars in galaxies are setting the primary property, instead of gas. This is largely due to the fact that obtaining large complete samples of galaxies with consistently measured gas properties is very challenging.  A first basic question is whether the correlation between halo mass and total baryon mass is stronger than the correlation with stellar mass.  This has been studied extensively with the Tully-Fisher relationship (e.g., \citealt{mcgaugh_etal:00}), and also with clustering of HI-selected samples (e.g., \citealt{guo_etal:17}).  At the current level of accuracy, the scatter between stellar mass--halo mass and baryon mass--halo mass appear to be comparable, although this may be more likely to break down for smaller mass galaxies which have high gas fractions \citep{bradford_etal:15}. Future large HI surveys should allow a more comprehensive study of this question.

An alternative galaxy property is galaxy size. Accurate statistical modeling is becoming especially interesting now that the relationship between galaxy sizes and stellar masses have been measured for larger samples in the local Universe \citep{cappellari_etal:13} and over a wide range of epochs with HST \citep[][]{vanderwel_etal:14}. Despite the complexity of galaxy formation, the tight scaling relations observed between various galaxy structural parameters imply a fairly tight connection between galaxy sizes and the properties of their dark  matter halos. We discuss joint modeling of two galaxy properties further in \S \ref{sec:complications_secondary}.

\begin{figure}[h]
\includegraphics[width=5in]{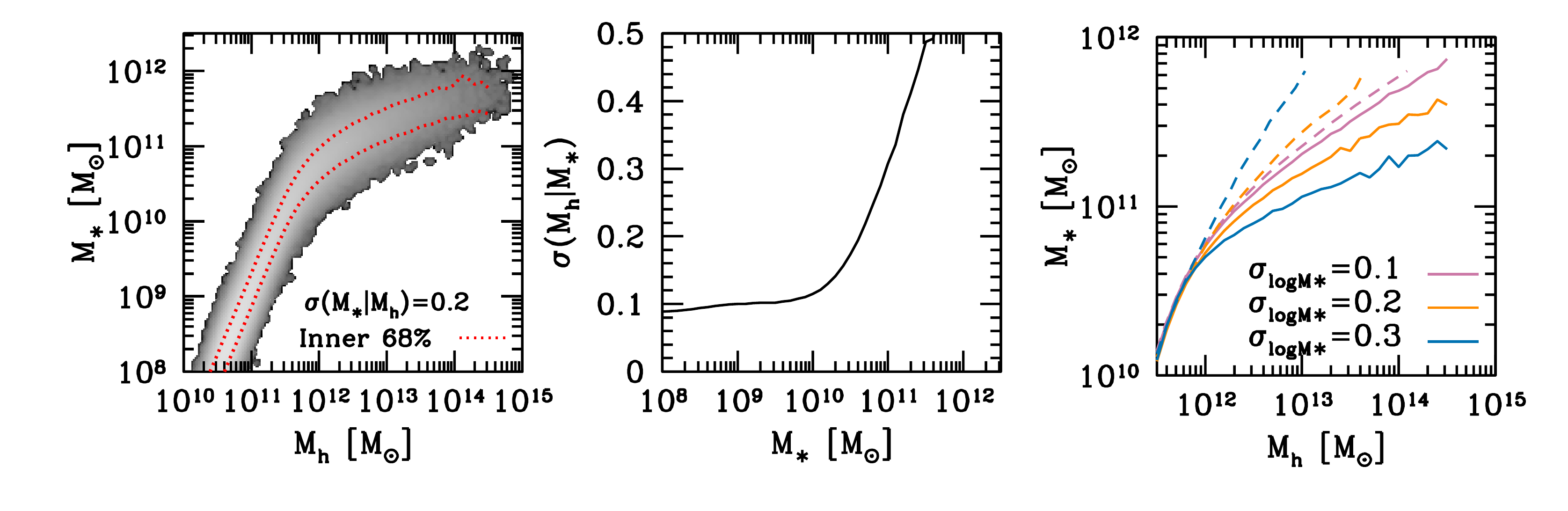}
\caption{\footnotesize{}{\it Left Panel:} The full distribution of halo and galaxy stellar masses for the abundance matching model from \textbf{Figure \ref{pedagogical1}} (i.e., the orange curve in that Figure). Both halos and subhalos are included.  The gray scale indicates the log of the number of objects in each bin in the two-dimensional plane. In this model, the scatter of stellar mass at fixed halo mass is constant, with $\slogm=0.2$ dex. The red dashed lines show the inner 68\% range of $\mgal$ at fixed $\mhalo$, which is 0.4 dex across the $y$-axis. However, due to the change in the slope of the SHMR at $\mhalo>10^{12}$ $\msol$, the distribution in $\mhalo$ at fixed $\mgal$ widens considerably above this mass. {\it Middle Panel}: The scatter in $\mhalo$ at fixed $\mgal$, $\sigma(\mhalo|\mgal)$ for the model in the left-hand panel. At low masses, this scatter is a constant value equal to roughly half of $\slogm$. Above the pivot point in the SHMR, the scatter monotonically increases, due both to the change in slope in the SHMR and to the changing slope of the halo mass function, which exponentially declines at high $\mhalo$. {\it Right Panel:} The solid curves show three different fits to the SDSS stellar mass function in \textbf{Figure \ref{pedagogical1}}, all with different values of $\slogm$. These solid curves all show the SHMR as the mean value of $\mgal$ in bins of $\mhalo$. The dashed curves show the corresponding reverse relationship, the mean value of $\mhalo$ in bins of $\mgal$. As $\slogm$ increases, the mean halo mass at fixed $\mgal$ decreases, even though the mean $\mgal$ at fixed $\mhalo$ decreases with increasing $\slogm$.}
\label{scatter_pedagogical}
\end{figure}

\subsection{Scatter}
\label{sec:scatter}
One of the principle questions about the galaxy--halo connection is how much scatter there is between the properties of galaxies at a given halo mass.  When comparing models, we consider scatter in the SHMR of central galaxies, e.g. the scatter in $\mgal$ at fixed $\mhalo$, which we refer to as $\slogm$. In abundance matching models the actual parameterization is in terms of the matching proxy (see the sidebar titled Different Halo Properties for Abundance Matching), and this scatter is generally assumed to be lognormal and constant across all halo masses.  This is both for convenience and because these assumptions appear to be consistent with all present data, including the results for fitting HOD models to galaxy threshold samples. See, for example, \cite{behroozi_etal:10} for the methodology of incorporating scatter in abundance matching while preserving the same stellar mass function.

HOD models parameterize the mean number of galaxies per halo.  For HOD models of luminosity or stellar mass threshold samples, the scatter is incorporated in the shape of the central galaxy occupation function. For central galaxies, because there can only be one or zero centrals, this translates to the probability of having a galaxy above the threshold as a function of halo mass. Thus the scatter derived in HOD models of this type is related to (but not {\it exactly}) the scatter in halo mass at a fixed galaxy mass (or luminosity). The HOD analysis of SDSS galaxies by \citet{zehavi_etal:11} found that this scatter---the scatter in $\mhalo$ at fixed $L_{\rm gal}$---monotonically rises with increasing $L_{\rm gal}$. 

\textbf{Figure \ref{scatter_pedagogical}} (left panel) shows the full two-dimensional distribution of halo masses and galaxy masses for the fiducial model that matches the stellar mass function and clustering of SDSS galaxies presented in \textbf{Figure \ref{pedagogical1}}. 
One can see that in such a model the scatter in $\mhalo$ at fixed $\mgal$ will depend sensitively on the value of $\mgal$ itself. The middle panel shows $\sigma(\mhalo|\mgal)$ explicitly. 
Above the pivot point the mean relation between  $\mgal$ and $\mhalo$ is shallower, and galaxies across a fixed width in $\log\mgal$ are spread out over a wider range of $\log\mhalo$. 

In abundance matching models, various choices for $\slogm$ can yield the same stellar mass function, but they do not predict the same spatial distribution of galaxies.  The right-hand panel of \textbf{Figure \ref{scatter_pedagogical}} shows three different SHMRs, all three of which provide good fits to the same SDSS stellar mass function but have three different values of $\slogm$. The solid curves show each SHMR as the mean $\mgal$ as a function of $\mhalo$. As $\slogm$ increases, the solid curves decrease at fixed $\mhalo$. In other words, the halo mass that yields a mean galaxy mass of $\meanmgal=10^{11.5}$ $\msol$ increases with $\slogm$. But recall that each model produces the same abundance of galaxies. Higher mass halos are less abundant, but the larger scatter brings in more low-mass halos that contain $10^{11.5}$ $\msol$ galaxies to preserve the same abundance. This has a direct impact on the clustering of galaxies, as higher mass halos above $\mhalo>10^{12}$ $\msol$ become increasingly more clustered (more highly biased). Thus the highest sensitivity to $\slogm$ from clustering measurements is in samples of galaxies at high masses or luminosities.

\begin{figure}
\includegraphics[width=5in]{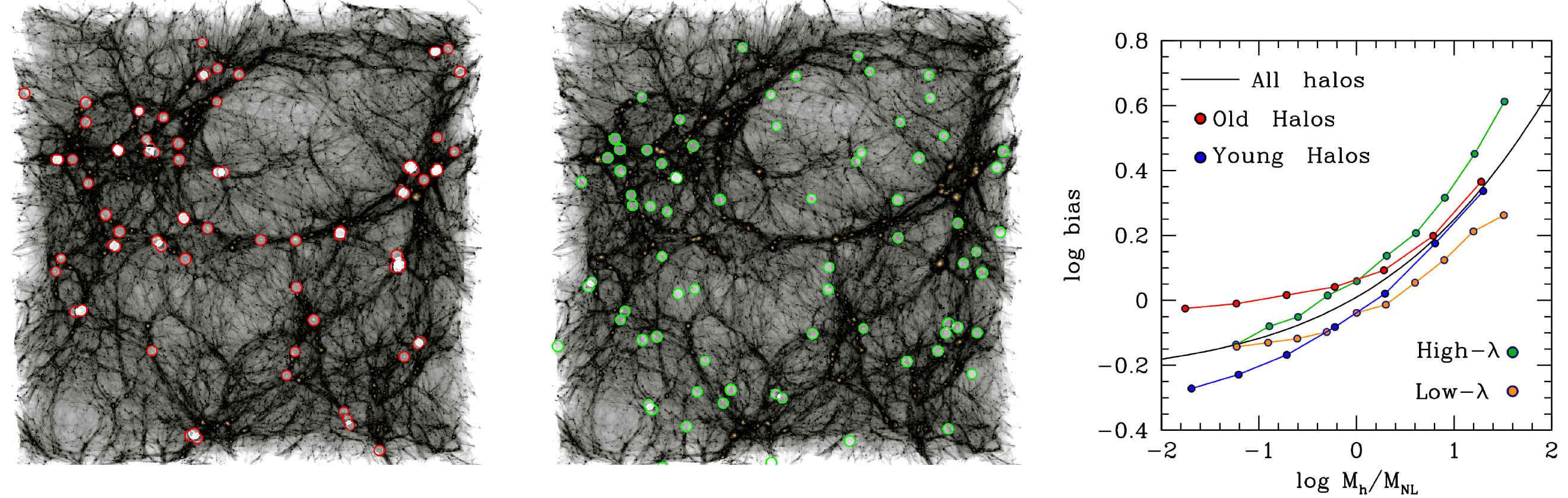}
\caption{Halo assembly bias, manifesting in concentration, halo formation time, and halo angular momentum. {\it Left Panel:} The gray scale shows the distribution of dark matter in a 90x90x30 Mpc slice of a cosmological simulation at $z=0$. The open red circles indicate the 5\% of halos at $\log M_h=10.8$ with the highest concentration. {\it Middle Panel:} The same slice is shown once again, but here the green circles show the locations of the 5\% of halos with the lowest concentration from the same halo mass range. {\it Right Panel:} The dependence of halo bias on secondary parameters. Bias here refers to clustering amplitude relative to dark matter, as defined in \S 3.2. The solid black curve shows the overall bias of dark matter halos as a function of halo mass at $z=0$. The red and blue points show the clustering for the 25\% of halos with the highest formation redshift and lowest formation redshift, respectively. Data are taken from \cite{li_etal:08}. The orange and green points show the clustering for the 20\% of halos with the highest and lowest angular momentum, respectively. Data are taken from \cite{bett_etal:07}. }
\label{assbias}
\end{figure}


\subsection{Assembly bias}
\label{sec:assbias}

Halos and galaxies experience a wide variety of assembly histories, even at fixed masses. Different assembly histories can influence the secondary properties of halos and galaxies. Assembly history also correlates with large-scale environment, yielding a correlation between some secondary properties and the spatial distribution of objects. In this subsection, we review how this assembly bias manifests for halos, and how this might propagate into the galaxy population.

\subsubsection{Halo assembly bias} 
Halo assembly bias refers to the effect that the clustering of halos at fixed mass has a dependence on properties other than $\mhalo$.
This dependence was initially detected when sorting halos at a given mass by formation time, concentration, and spin \citep{wechsler:01,gao_etal:05, wechsler_etal:06}, as well as for simulated galaxy samples \citep{croton_etal:07}, and has now been studied extensively in the literature for a number of properties of dark matter halos (see, e.g., \citealt{mao_etal:17} for a recent comprehensive study). \textbf{Figure \ref{assbias}} shows an example of halo assembly bias that can be seen visually in the distribution of low-mass halos, as well as a quantitative assessment. Different assembly bias effects can be created from tidal forces in the density field (\cite{hahn_etal:09}), and from the statistics of peaks in a Gaussian random field (\cite{dalal_etal:08}). 

Although its existence for dark matter halos is now well established, there are many open questions about the assembly bias signal for various halo properties and whether or to what extent this effect propagates into the clustering of galaxies.   We summarize some of these properties in the sidebar titled What Halo Properties Show Secondary Bias?, but note that in general, the secondary bias of various halo properties can be complex, and can have different mass and redshift dependence \citep{salcedo_etal:17}. Furthermore, \cite{mao_etal:17} have shown that even properties that are highly correlated do not necessarily have the same clustering signal. This may make first principles predictions for the expected galaxy assembly bias challenging, but it also indicates that precision measurements of galaxy clustering may provide insights into complex details of structure formation and the dependence of galaxy properties on halo properties.

\begin{marginnote}[]
\entry{Halo concentration}{Halo concentration is defined as $c\equiv R_{\rm halo}/r_s$, where $r_s$ is the scale at which the logarithmic slope of the internal density profile is $-2$.}
\end{marginnote}

\begin{textbox}
\subsection{Important Definitions}
\begin{itemize}
\item Bias: The clustering of a set of objects relative to the clustering of dark matter.
\item Halo Assembly Bias or Secondary Bias: At fixed halo mass, the clustering of dark matter halos depends on secondary halo properties
(which are generally correlated with the assembly history of the dark matter halo).
\item Galaxy Assembly Bias: At fixed halo mass, the galaxy properties or number of galaxies within dark matter halos may depend on secondary halo properties that themselves show a halo assembly bias signature. This includes both central and satellite galaxies. {\em Note that this definition, proposed by \cite{zentner_etal:14} is not exactly parallel with the definition of halo assembly bias definition, and is dependent on the existence of halo assembly bias.} It is well known that galaxy clustering depends on secondary galaxy properties, but this can be caused by different satellite fractions or a dependence on halo mass for the secondary property.

\end{itemize}
\subsubsection{What halo properties show secondary bias?}
\begin{itemize}
\item Halo properties that show stronger assembly bias at lower masses: half-mass redshift $\zhalf$, formation redshift $a_c$, fractional halo growth rate $\Delta M/M$. These quantities are usually defined with respect to $\mhalo$ at $z=0$, but they show assembly bias when defined at any redshift.
\item Halo properties that show stronger assembly bias at higher masses: angular momentum $\lambda$, amount of substructure.
\item Halo concentration shows signals at high and low masses, but the sign of the signal switches at the non-linear halo mass $M\sim \mnl$.
\end{itemize}
\end{textbox}

\subsubsection{Theoretical models of galaxy assembly bias}

In the abundance matching paradigm, the relation between galaxies and halos is set by the halo property that is used to match to the given galaxy property. However, there is no {\em a priori} reason to limit the model to a single halo property. In the literature, there have been multiple models for multi-parameter galaxy--halo connections. Here we list two that vary the galaxy clustering properties for luminosity or mass selected galaxy samples; models that include assembly bias for secondary properties like color or star formation rate are discussed in \S \ref{sec:complications_secondary}.

{\bf Composite abundance matching:} In this model, the abundance matching parameter is a combination of two or more halo properties. In the model by \citet{Lehmann17}, the abundance matching parameter is

\begin{equation}
v_\alpha = v_{\rm vir}\left(\frac{v_{\rm max}}{v_{\rm vir}}\right)^\alpha,
\end{equation}

\noindent where $v_{\rm vir}=(GM_{\rm vir}/R_{\rm vir})^{1/2}$ is the virial velocity of the halo, and $v_{\rm max}$ is the maximum circular velocity within the halo, and $\alpha$ is a free parameter. When $\alpha=0$, the abundance matching parameter is $v_{\rm vir}$, which is directly proportional to $M_{\rm vir}^{1/3}$---i.e., abundance matching galaxy mass to halo mass. When $\alpha=1$, the abundance matching parameter is $v_{\rm max}$, which is directly related to halo concentration. This process is implemented on halo catalogs that resolve substructure. In \citet{Lehmann17} these halo quantities are evaluated at the epoch of peak halo mass. Increasing $\alpha$, therefore, has the effect of increasing the clustering at fixed $\mgal$ (because of the effect of concentration on clustering, as shown in \textbf{Figure \ref{assbias}}). This freedom essentially parameterizes our current ignorance about how much galaxy properties depend on concentration or formation time at a given halo mass, which impacts both the assembly bias and the satellite fraction at a given stellar mass.

\begin{marginnote}[]
\entry{Non-linear halo mass, ($\mnl$)}{This is defined as the mass at which the rms matter fluctuations on the Lagrangian scale of the halo are equal to $\delta_{\rm crit}=1.686$ in linear theory. $\mnl\sim 10^{12.8}$ $\msol$ at $z=0$ in the Planck cosmology, and decreases with increasing redshift.
}
\end{marginnote}

{\bf Modified halo occupation models:} The previous approach requires simulations that resolve substructure. An alternative is to take parameterized HOD models---i.e., models in which the mean occupation function of central and satellite galaxies are parameterized as a function of $\mhalo$---and modify them to include additional halo properties in the mean occupation function. These models require cosmological simulations or a multi-parameter emulator based on such simulations, but no longer require that substructure be tracked. \citet{hearin_etal:16} proposed {\em decorated HODs}, with the secondary parameter being halo concentration;  \citet{tinker_etal:08_voids} used large-scale density as the secondary parameter to make models to compare to measurements of clustering and voids. In the decorated HOD models, the mean occupation function is increased or decreased, relative to the mean at a given value of $\mhalo$, on the basis of whether the concentration is above or below the median $c$. The change in the HOD can be either a step function or a smooth function in $c$, and can be either increasing with $c$ or decreasing with $c$. Concentration is also fungible with other halo properties, but estimating $c$ for a halo has the advantage that it does not require full halo merger trees, as $\zhalf$, $a_c$, and various other properties do. Given the strong correlation between $c$ and halo growth quantities---at least at mass scales below the cluster scale---the results seemed likely to be similar; however, the work of \citet{mao_etal:17} urges caution in making these simplifying assumptions.

\subsection{Centrals vs. Satellites}
\label{sec:complications_censat}

One of the keys to creating a robust model of the spatial distribution of galaxies is getting the correct fraction of satellite galaxies as a function of galaxy mass.  One of the strongest arguments in favor of the galaxy--halo connection as a model with physical underpinnings is that a simple abundance matching model generally reproduces the small-scale clustering as a function of galaxy luminosity or mass, as well as its redshift dependence.  But, as \textbf{Figure \ref{reddick13}} demonstrates, details of the relative treatment of central and satellite galaxies matter when compared to the high precision of galaxy clustering measurements now available at many epochs.

It is not a requirement of abundance matching or SHMR models that they treat the galaxies within host halos and subhalos the same. Indeed, the $M_{0,\rm peak}$ and $V_{0,\rm peak}$ models in that figure explicitly separate the two.  Even for models that are matched to halos in the same way for centrals and satellites, the details can change the stellar mass of centrals vs. satellites at a given halo mass; see for example the change as one varies $\alpha$ in Fig. 5 of \cite{Lehmann17}.  In the SHMR model of \cite{leauthaud_etal:11a,leauthaud_etal:12}, this flexibility is introduced explicitly by having the SHMR only apply to central galaxies, and having the number of satellites within a halo vary freely, independently of the number of subhalos in the halo. Alternatively, one can have two different SHMR functions---one for centrals and one for satellites---as done in \cite{watson_conroy:13}. Although the details do matter to within the precision of SDSS-like clustering measurements, \cite{watson_conroy:13} concluded that in such a framework, the preferred model was one in which the SHMR for centrals and satellites was nearly the same.

\subsection{Occupation properties of satellite galaxies}
\label{sec:poisson}

For models that can calculate clustering without populating a simulation---and this includes HOD, CLF, and some parameterized SHMR models \citep{vale_ostriker:04,leauthaud_etal:11a}---assumptions are nearly always made that should not be taken as gospel in this era of precision measurements from large galaxy redshift surveys. The first assumption is that satellite galaxies are assumed to follow the Navarro–Frenk–White (NFW) density profile, which may not be true at very small scales \citep{watson_etal:12}. The second assumption is that the second moment of the probability distribution of satellite galaxy occupation has been assumed to be Poisson in most previous work.

Models for the halo occupation distribution require an assumption about both the mean and moments of the occupation distribution.  Higher order moments can have an impact on clustering properties and on cosmological systematics, so it is important to know how robust the simplifying assumption of a Poisson distribution is. The scatter in the number of subhalos at fixed mass was shown to be super-Poissonian for small average occupation numbers \citep{boylan_kolchin_etal:10,busha_etal:11}; \cite{wu_etal:13} showed that the scatter can in detail depend on how galaxies or subhalos are selected. \cite{mao_etal:15} showed that this can be explained by Poisson scatter at fixed mass {\em and formation history}, combined with dependence of the number of subhalos on other properties (e.g. formation history, environment, or concentration) at fixed mass; \cite{jiang_vdb:17} claimed that this is not quite true, and that the distribution is both sub-Poissonian at small average occupation number and super-Poissonian at large average occupation number.  They present an accurate fitting function for the distribution and show how this can impact clustering.

\subsection{Joint modeling of mass and secondary properties}
\label{sec:complications_secondary}
Thus far we have discussed the relationship between a single galaxy property (mass or luminosity) and a single or composite halo property.  Of course, the full galaxy--halo connection is more complicated, and for a full description of galaxy formation we will be interested in the full multivariate connection between galaxy and halo properties over time.  Here we review approaches to modeling a secondary property in addition to stellar mass or luminosity.

Halo occupation methods can be extended to incorporate more than one galaxy property. For example, \cite{skibba_sheth:09} developed an HOD model that incorporated galaxy color. \cite{xu_etal:18_ccmd} recently created a conditional color-magnitude distribution, an extension of the CLF methodology. Both of these methods parameterize the mean occupation of galaxies within halos using $\mhalo$ only.

A complementary way to incorporate secondary properties into the galaxy--halo connection is through  `conditional abundance matching', as proposed by \cite{hearin_watson:13}. In this framework, halo mass is first abundance matched to $\mgal$ or luminosity, and then secondary galaxy properties are abundance matched to secondary halo properties in narrow bins of $\mgal$ or $\mhalo$. This framework was explored and refined using galaxy color \citep{hearin_watson:13,hearin_etal:14} and star formation rate \citep{watson_etal:15} as the secondary galaxy property. For the secondary halo property, variations on halo age and formation time were employed. At fixed halo mass, the earliest forming halos contained galaxies with the reddest colors or lowest star formation rates, thus this model can be described as an `age-matching model.' However, this framework extends naturally to any secondary galaxy property or halo property. For example, \cite{tinker_etal:17_p3} used conditional abundance matching to connect galaxy morphology to halo angular momentum at fixed $\mhalo$. \cite{hearin_etal:17} connected galaxy size to halo virial radius at the time of $M_{\rm peak}$ through a related approach (we discuss constraints on galaxy sizes in \S \ref{sec:constraints_sizes}).

Conditional abundance matching is an ideal tool for comparing models of galaxy assembly bias to observational data. If the secondary halo property used exhibits an assembly bias signal (such as age, concentration, or spin), then this would manifest as a clustering dependence in the secondary galaxy property.

Several recent studies have begun to explore the evolution of gas in galaxies in more detail in both hydrodynamical simulations \citep{power_etal:10,bahe_etal:15,lagos_etal:14} and semi-analytic models \citep{popping_etal:14,lu_etal:14}.
Empirical models for full galaxy populations are starting to be used to model gas as well.  Recently, \cite{popping_etal:15} used an empirical model for the star formation histories along with observed scaling relationships between stellar and gas densities to develop a model which traces the evolution of gas properties over time.  We expect modeling the gas--halo connection to be an area of extensive future work, aided by new insight from UV absorption studies, X-ray studies, HI surveys, and  Sunyaev-Zel'dovich (SZ) surveys to develop comprehensive models for the evolution of gas properties that should substantially impact our understanding of galaxy formation physics.

\section{CURRENT CONSTRAINTS ON THE GALAXY--HALO CONNECTION AS EXPRESSED BY THE SHMR}
\label{sec:constraints}
In this section, we discuss constraints on different aspects of the galaxy--halo connection as expressed by the SHMR. The mean relation is discussed in \S 5.1, the scatter in this relation in \S 5.2, and the evolution of the mean relation in \S 5.3. \S 5.4 discusses observational tests of whether halo properties other than mass influence the stellar mass or luminosity of galaxies. This section closes with a brief overview of systematic uncertainties, both observational and theoretical, that need to be taken into account when interpreting current constraints. Section \ref{sec:constraints_secondary} considers second-parameter connections involving additional parameters beyond mass.

\subsection{The mean stellar-to-halo mass relation for central galaxies}
Figure \ref{pretty_gmhmr} presented the constraints on the SHMR in the local universe, painting a consistent picture of the relationship between halos and total galaxy mass from a variety of methods and datasets. These approaches have now been described in the preceding sections.  There is remarkable consistency on the general outlines of the SHMR from these methods, including constraints from non-parametric abundance matching \citep{behroozi_etal:10,reddick_etal:13,behroozi_etal:13}, the parameterized SHMR inferred from abundance matching \citep{guo_etal:10, moster_etal:13,moster_etal:17}, the halo occupation distribution \citep{zheng_etal:07}, the conditional luminosity function \citep{yang_etal:12}, abundance matching from X-ray clusters \citep{kravtsov_etal:14}, or models based on evolving galaxies within their dark matter halo histories constrained by galaxy clustering and galaxy-galaxy lensing \citep{behroozi_etal:17}.  

These approaches use different observables and different modeling techniques.  For example, the HOD and CLF leave satellite occupation as a free parameter to be constrained by the data, whereas abundance matching techniques are constrained to match satellites onto subhalos within simulations using a global galaxy--halo relation. Observables used to constrain these approaches include the stellar mass function at a given epoch, the stellar mass and star formation rates as a function of time, the group stellar mass function and conditional stellar mass function of galaxies in groups, galaxy clustering, galaxy--galaxy lensing, and satellite kinematics.  The biggest discrepancy between these methods in terms of the mean relation is due to systematics in the measurement of the stellar mass itself at the highest mass end.  At the lowest mass end, uncertainties are dominated by the fact that galaxy samples are still small and likely incomplete. We discuss these issues further in \S  \ref{sec:systematic_obs}.

\subsection{Scatter for central galaxies}
One of the most important aspects of the galaxy--halo connection is the scatter in central galaxy stellar mass at fixed halo mass,  $\slogm$. It is not possible to constrain $\slogm$ from the abundance of galaxies alone;  here we present a number of possibilities that can constrain this parameter: galaxy clustering, galaxy groups and clusters, satellite kinematics, galaxy--galaxy lensing, and galaxy scaling relations.

\begin{figure}[t]
\includegraphics[width=5in]{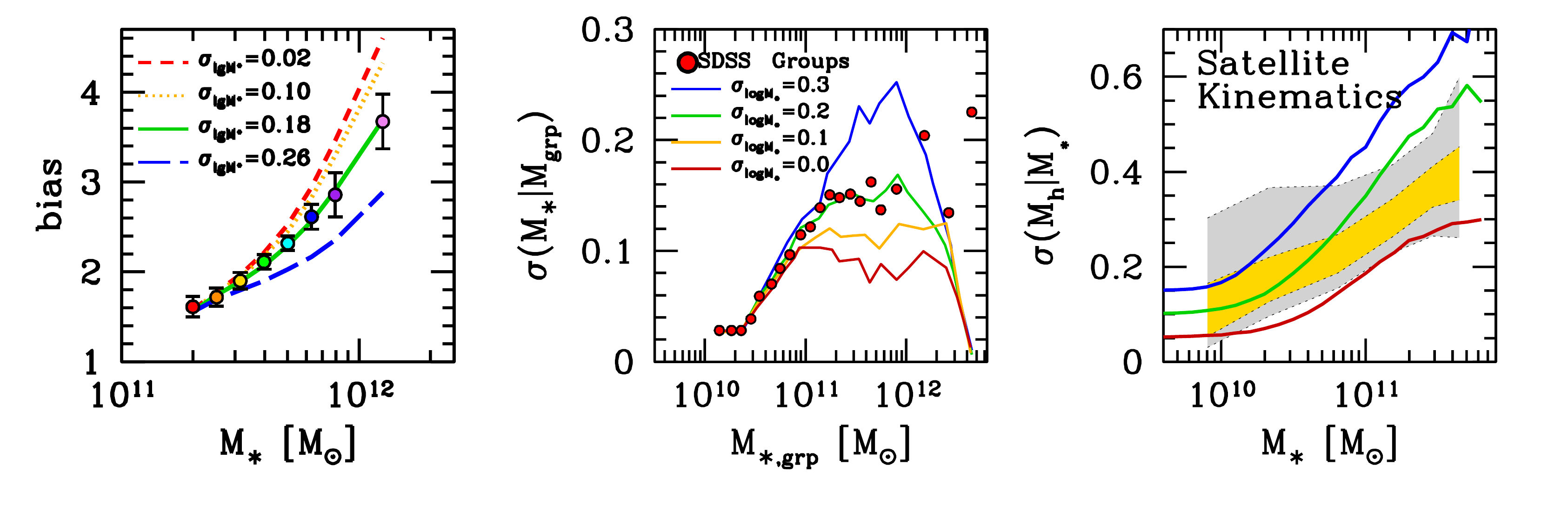}
\caption{{\it Left Panel:} Points with errors show bias as a function of $\mgal$ for $z=0.5$ galaxies in the BOSS survey from \citet{tinker_etal:17_boss}. Curves indicate SHMR fits to the BOSS stellar mass function with various values of $\slogm$. {\it Middle Panel:} Points show the scatter in $\mgal$ at fixed group stellar mass for central galaxies in $z\approx 0$ SDSS group catalogs. Curves show predictions from abundance matching models with different values of $\slogm$. All theoretical models have been run through the group finding algorithm to create an apples-to-apples comparison. All data and models taken from \citet{reddick_etal:13}. The three data points at $\log M_{\ast \rm grp}>12$ are not significant owing to low numbers of groups above this mass scale. {\it Right Panel:} The scatter in $\mhalo$ at fixed $\mgal$, derived from analysis of satellite kinematics by \citet{more_etal:11}.  Contours indicate 68\% (yellow) and 95\% (grey) confidence intervals. 
The solid curves show predictions for the quantity using the three abundance matching models in \textbf{Figure \ref{scatter_pedagogical}}, in which $\slogm=0.1$, 0.2, and 0.3.}
\label{scatter_3win}
\end{figure}

\textbf{Galaxy clustering:} Halo bias is a strong function of mass for halos above $\mnl$, so in this regime galaxy clustering can provide a strong constraint on scatter.  The left panel of \textbf{Figure \ref{scatter_3win}} gives an example, based on galaxy clustering at $z\sim 0.5$ from the BOSS survey \citep{tinker_etal:17_boss}.  Different curves represent the predictions of SHMR models with different scatter values, but only $\slogm=0.18$ provides a good fit to the observed bias of BOSS galaxies. The constraining power of clustering is primarily at the massive end; this result is best characterized as a constraint in $\slogm$ in the halo mass range of $\log\mhalo=[12.7,13.7]$.  Using a combination of galaxy clustering and galaxy lensing of $z=0$ SDSS galaxies, an independent study by \citet{zu_mandelbaum:15} obtained a constraint on the scatter at $\mhalo=10^{12}$ $\msol$ ($\slogm=0.22^{+0.02}_{-0.01}$) and at $\mhalo=10^{14}$ $\msol$ ($\slogm=0.18\pm 0.01$).

\textbf{Galaxy groups and clusters:} Galaxy groups and clusters can constrain the scatter by directly measuring the properties of central galaxies in groups of clusters of a given mass.  Using a sample of galaxy clusters with well-measured X-ray halo mass estimates, \citet{kravtsov_etal:14} derived a value of $\slogm=0.17$ by directly measuring the scatter around the mean $\mgal$-$\mhalo$ relation. This measurement probes the halo mass range of $\log\mhalo=[14.0,15.0]$. Similar measurements can be made by galaxy group catalogs; here a critical issue is how well the mass of the group can be measured.  The middle panel of \textbf{Figure \ref{scatter_3win}} shows scatter in the mass of the central group galaxy at fixed group halo mass, $\sigma(M_{\rm cen}|M_{\rm grp})$ from \cite{reddick_etal:13}.  To facilitate a proper comparison with theory, \citet{reddick_etal:13} created abundance matching mocks constrained to the SDSS stellar mass function, which were run through the group finder and processed in the same manner as the data. The figure shows the mock values of $\sigma(M_{\rm cen}|M_{\rm grp})$ for four different values of $\slogm$. Data at low stellar masses have no constraining power because the masses of these groups are not reliable, but using data at $\mgal\simeq 10^{11}$ $\msol$, \citet{reddick_etal:13} determine a best-fit value of $\slogm=0.20\pm 0.02$ (note that this value refers to the scatter in central galaxy properties at fixed halo mass). This constraint likely has modest sensitivity to the abundance matching proxy used, because the measured scatter at fixed group mass depends on the correlation between central galaxy mass and total group mass. Future surveys, for example, the DESI Bright Galaxy Survey \citep{desi}, may be able push this scatter constraint down to lower masses because lower mass groups will contain significantly more galaxies.

\textbf{Satellite kinematics:} Satellite kinematics offer a complementary approach to constraining scatter by probing the dark matter gravitational potential around central galaxies. With a sufficiently deep spectroscopic sample, satellite kinematics can constrain $\slogm$ around low-mass galaxies and halos below the regime where clustering loses its sensitivity.  The right-hand panel of \textbf{Figure \ref{scatter_3win}} shows constraints on scatter using kinematics of satellite galaxies from \citet{more_etal:11}. In this analysis, the scatter is presented as $\sigma(\mhalo|\mgal)$. Solid curves plotted over the contours show $\sigma(\mhalo|\mgal)$ for the three SHMR fits to the SDSS stellar mass function from \textbf{Figure \ref{scatter_pedagogical}}: $\slogm=0.1$, 0.2, and 0.3.  We note that these comparisons between theory and data are not precise because they are based on different assumptions about the stellar mass measurements.  However, these data are consistent with a model in which $\slogm=0.2$ or somewhat smaller, which is consistent with constraints from clustering, lensing, and group statistics.

\textbf{Scaling relations:} Before the advent of surveys that could measure clustering and lensing for large galaxy samples, the observation of tight dynamical scaling relations was the best clue that galaxy and halo properties were tightly connected. These have continued to provide interesting constraints on galaxy formation models \citep{governato_etal:07,somerville:08}; for example simultaneously matching clustering, abundance, and the Tully-Fisher relation has been challenging.  Recently, studies have begun to combine these constraints to test empirical models \citep{desmond_wechsler:15,desmond_wechsler:17}. Although with current samples these are not yet competitive with clustering to constrain scatter in the primary galaxy parameter (stellar mass or luminosity) at mass above $\sim 10^{12} \msun$, they can provide useful constraints at lower masses and also can constrain the covariate scatter between e.g. mass and size at fixed halo mass.

\textbf{Measurement error:} 
We note that most observational constraints on $\slogm$ represent the quadratic sum of intrinsic scatter and measurement scatter. The uncertainties in estimating stellar mass are much discussed (e.g., \citealt{conroy_etal:09}), and typical uncertainties in $\mgal$ range from 0.2--0.3 dex (\citealt{mobasher_etal:15}). However, these uncertainties represent a combination of scatter and overall biases in stellar mass estimates. While the former contributes to $\slogm$, the latter does not. \cite{tinker_etal:17_boss} estimated a lower limit to the observational scatter to be 0.11 dex for their stellar masses, yielding an upper limit to the intrinsic scatter of 0.16 dex. Further understanding of measurement scatter will enhance our ability to constrain the intrinsic $\slogm$.

\subsubsection{Comparison with  galaxy formation models}

\begin{figure}[t]
\includegraphics[width=5in]{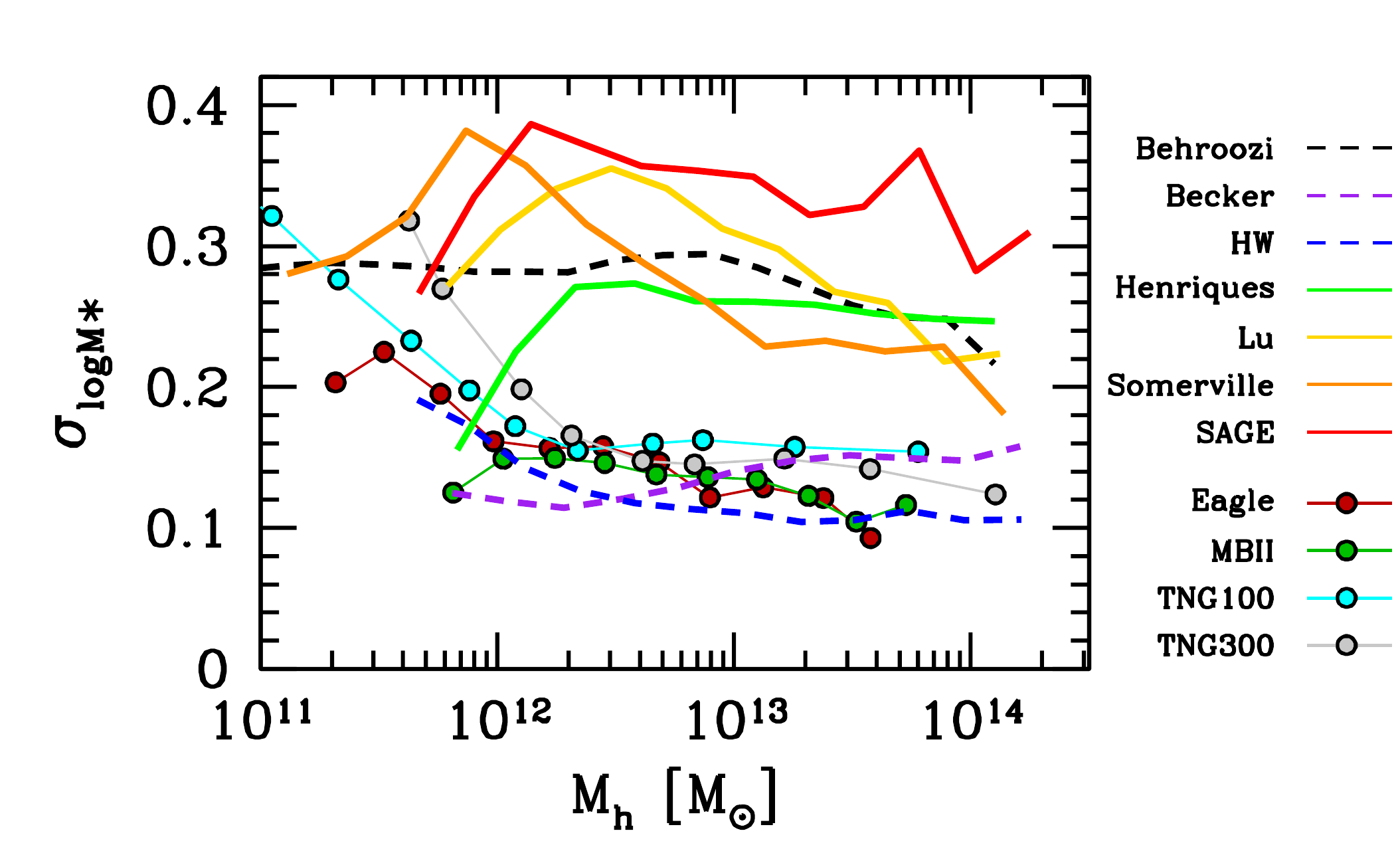}
\caption{Scatter in the stellar mass of central galaxies at a given halo mass from various theoretical and empirical models. Models include four hydrodynamical simulations, Massive Black II \citep{khandai_etal:15}, Eagle \citep{eagle} and two IllustrisTNG simulations \citep{pillepich_etal:18} (colored circles connected by thin solid lines); four semi-analytic models, from \cite{henriques_etal:15},  \cite{lu_etal:14}, \cite{somerville_etal:12}, and the SAGE model from {\tt https://tao.asvo.org.au/tao/}  (thick solid lines); and three empirical models, from \citet{behroozi_etal:17}, \citet{becker:15}, and \citet{hearin_watson:13} (dashed lines).
}
\label{scatter_theory}
\end{figure}

How do these constraints compare to the scatter we expect from physical models of galaxy formation?  \textbf{Figure \ref{scatter_theory}} shows $\slogm$ as a function of $\mhalo$ for a range of models, including cosmological hydrodynamic simulations of galaxy formation, semi-analytic galaxy formation models, and empirical models.  As discussed in the previous section and as shown in \textbf{Figure 7}, several recent measurements indicate that the scatter in stellar mass at fixed halo mass is quite well constrained to be below 0.2 dex at the high mass end, likely below 0.16 dex when considering only the intrinsic scatter predicted by these models.  This is in good agreement with current predictions from hydrodynamical simulations, as well as with some of the empirical models.  We note however that all of the semi-analytic models shown here, as well as the \cite{behroozi_etal:17} model which traces galaxy star formation through dark matter merging histories, have somewhat larger scatter at the high mass end.  This may be due to inadequate correlation between when galaxies are quenched and the properties of halos at that time; it will be important to understand what differences in the physical parameterizations lead to this difference.  

Thus far we have discussed primarily the regime above the pivot point around $M = 10^{12}$, where the scatter is well constrained by several observational measures.  As seen in \textbf{Figure \ref{scatter_3win}}, at present, clustering, group catalogs, and satellite kinematics only produce strong constraints above this regime, although the tightness of galaxy scaling relations provide some constraint at lower masses. 
At halo masses above $\sim 10^{12.5}\msun$ where the observational constraints are most robust, most models do predict roughly constant scatter.   At lower masses, most models predict increasing scatter, though care must be taken with resolution effects (for example, if the merger histories are measured with low resolution, this may create artificial scatter) and also with definitions --- which scatter is being considered?  When considering scatter at the lowest masses, down to the dwarf scale, we have very little direct observational information about the scatter of central galaxies, and many studies (both observational and theoretical) have so far considered primarily the scatter of satellites within the Milky Way or similar simulated galaxies.  Here then, it is important to distinguish scatter due to satellite stripping from scatter due to the formation processes for central galaxies. For example \cite{munshi_etal:17} claim to find high scatter in low mass galaxies, but in fact show rather small scatter, $\sigma < 0.25$ dex, for the pre-stripped quantities.  Overall, the evidence suggests that there is slightly more scatter in the stellar masses of dwarf galaxies; understanding how large this scatter is a critical piece of tests of the CDM model in this regime and will be a major area of future work as samples of dwarf galaxies increase \citep{bullock_boylankolchin:17}.

\subsection{Evolution}
Above we have primarily focused on the galaxy--halo connection in the local universe, where it is accurately constrained by the abundance and detailed clustering properties of galaxies.  How do we expect it to evolve?  Stellar mass functions have been measured up to $z\sim 8$, allowing abundance matching to be applied over the majority of the history of the universe.  The first study to investigate the evolution of the SHMR over most of cosmic time was performed by \cite{conroy_wechsler:09} who used abundance matching at individual epochs combined with information about halo accretion over time to infer the evolution of the SHMR and galaxy assembly histories.  This has since been extended by other authors using halo merger trees directly, as well as extensive updated information on the populations of high redshift galaxies.  

\begin{figure}
  \begin{minipage}[b]{0.45\textwidth}
    \includegraphics[width=\textwidth]{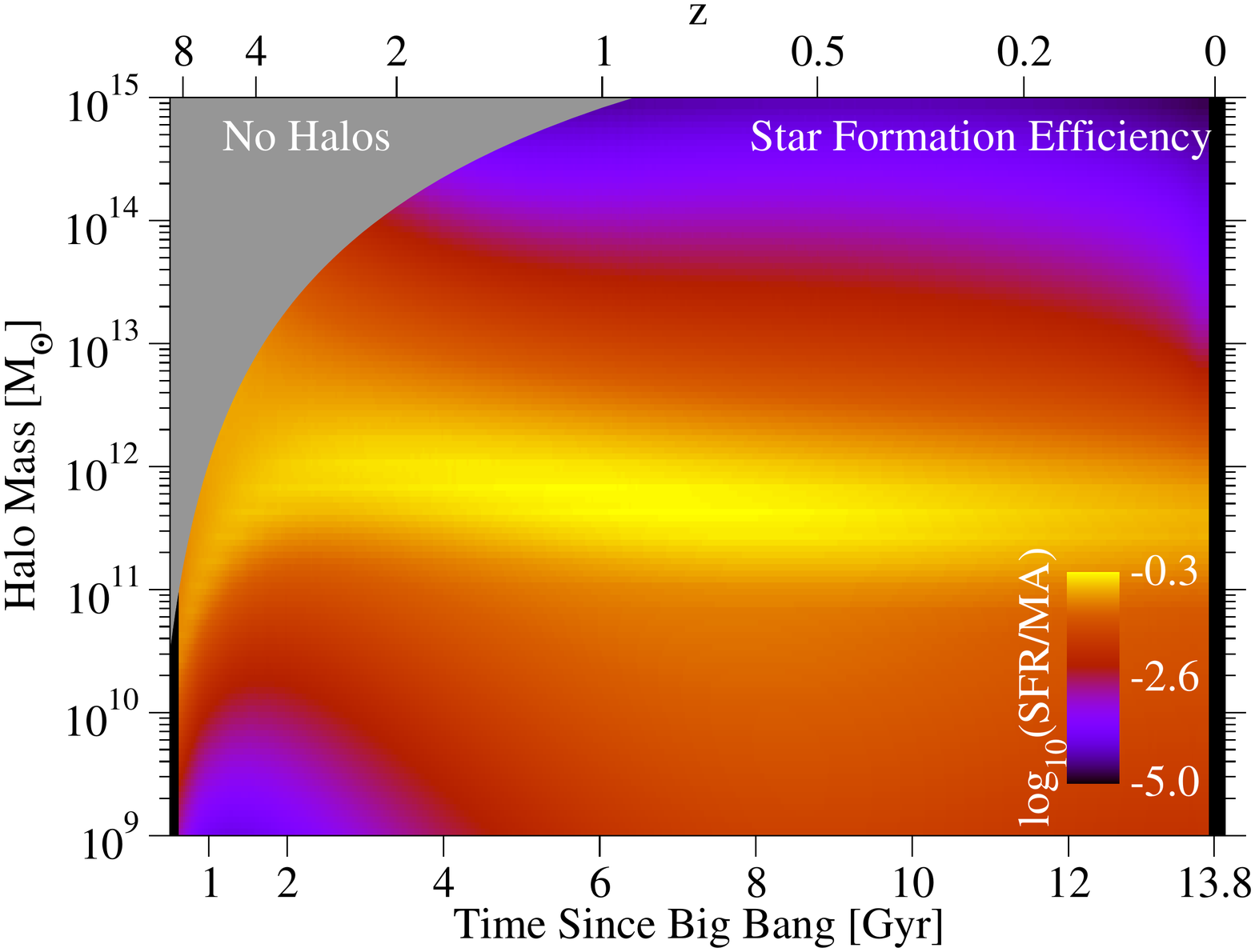}
  \end{minipage}
  \begin{minipage}[b]{0.45\textwidth}
    \includegraphics[width=\textwidth]{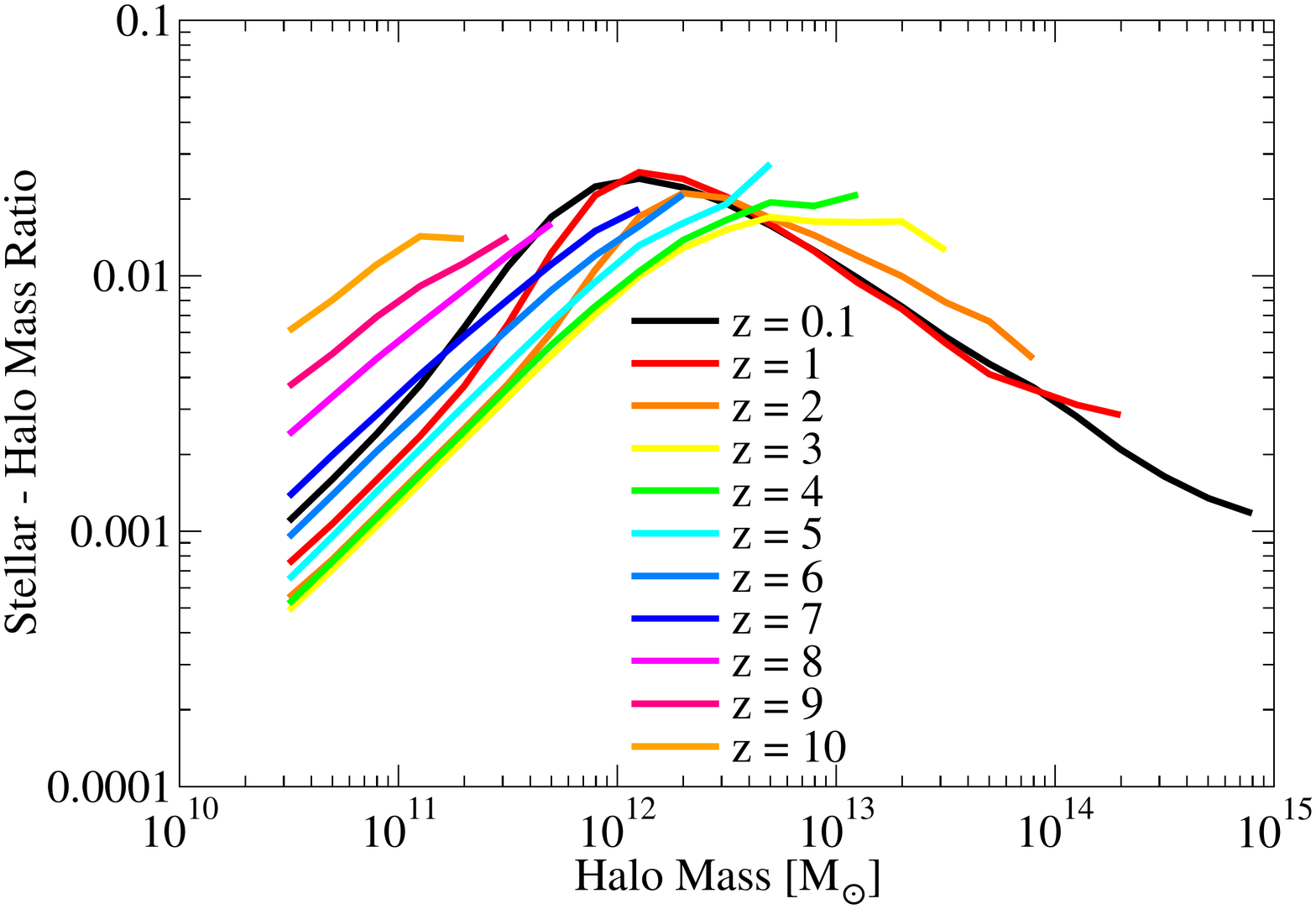}
  \end{minipage}
\caption{{\em Left:} Star formation efficiency (defined here as the star formation rate divided by the halo mass accretion rate) as a function of halo mass and redshift. Two thirds of all star formation occurs in within a factor of three of the peak halo masses.  From \cite{behroozi_etal:13_letter}.  {\em Right:} Evolution of the SHMR with redshift as inferred from the empirical model of \cite{behroozi_etal:17}.}
\label{evolution}
\end{figure}

The primary conclusion from a range of studies is that the star formation efficiency, defined here as the ratio of the star formation rate divided by the mass accretion rate, is a strong function of mass, peaking at roughly $10^{12} \msun$, but a very weak function of redshift.  This is shown in the left panel of \textbf{Figure \ref{evolution}}, based on the results of \cite{behroozi_etal:13_letter}, which synthesized a range of measurements. This study found that two-thirds of all star formation occurs in a relatively narrow range of halo masses.  We note that the halo mass accretion rate is declining with time, so the star formation rate itself is significantly higher for a galaxy at a given stellar mass at higher redshift. A typical galaxy that lives in a massive halo today started forming stars early, but then at relatively early times moved out of this efficient mass range.  A typical galaxy in a smaller halo started forming stars later, but spends a longer region in this regime of efficient star formation. The resulting SHMR as inferred by the study of \cite{behroozi_etal:17}  is shown in the right hand panel.  Somewhat surprisingly, the SHMR evolves rather little with time; the peak of the relation is nearly constant to $z \sim 3$.  Because low mass galaxies are still building up, their stellar mass to halo mass ratio increases over this time.  Above $z \sim 3$, nearly all galaxies are still forming stars efficiently, and it is unclear whether the SHMR turns over and declines at the highest masses.

Galaxy clustering and galaxy lensing can be used to test these models out to higher redshift. This has been done using clustering alone in surveys like BOSS, DEEP, and PRIMUS out to $z\sim1$ both for typical $\sim L_\ast$ galaxies \citep{zheng_etal:07,coil_etal:08, abbas_etal:10,tinker_wetzel:10,wake_etal:11,skibba_etal:15} and for massive galaxies \citep{wake_etal:08,zheng_etal:09,white_etal:11,guo_etal:14,zhai_etal:16}. Clustering and lensing were combined to constrain the SHMR in COSMOS out to $z\sim 1$ (\citealt{leauthaud_etal:12}). Lensing by itself can also provide constraints at higher redshift (\citealt{hudson_etal:15}), although these constraints can be significantly tightened by bringing in measurements of clustering and/or the stellar mass function (\citealt{vaniutert_etal:16}).  We expect that the current and next generation of large imaging and spectroscopic surveys will dramatically increase the statistical power of these constraints to higher redshift.

\subsection{How does assembly bias manifest observationally?}

In models of galaxy assembly bias, star formation efficiency may depend on halo properties other than halo mass. The consequence of such a secondary correlation would be that the residuals of the SHMR would correlate with other halo properties. Observationally, this could impact the clustering of galaxies as a function of stellar mass. Such an effect has been searched for by a number of authors. \citet{Lehmann17} and \citet{zentner_etal:16} have fit $z=0$ clustering measurements using models that include secondary halo properties. \citet{Lehmann17} used the composite abundance matching method, whereas \citet{zentner_etal:16} used the enhanced HOD method. Both models used halo concentration as their secondary halo property. Both found at least one luminosity threshold sample that preferred a model in which galaxy occupation depended on $c$ in such a way to increase the chance of finding a central galaxy in higher concentration halos, thus boosting the clustering of the model galaxies. This is in agreement with some results from both semi-analytic models and cosmological hydrodynamic simulations, which we present in \S 6. 

A significant caveat to both of these analyses is that they are performed at a fixed cosmology. The amplitude of galaxy clustering for a fixed HOD is sensitive to the amplitude of dark matter fluctuations, which can be influenced by $\sigma_8$, $\Omega_m$, and to a lesser extent other cosmological parameters. Thus it is an open question whether galaxy assembly bias is required to match the observed level of galaxy clustering over the entire range of cosmological parameter space currently allowed by cosmic microwave background and other large-scale structure probes. Alternatively, if one's goal is to constrain cosmology from analysis of non-linear clustering, galaxy assembly bias can be degenerate with changes in cosmological parameters. Thus, robust analyses of small-scale galaxy clustering must take galaxy assembly bias into account in order to yield robust cosmological constraints. 

\subsection{Systematic uncertainties}
\label{sec:systematic}
As the statistical power of the data constraining the galaxy--halo connection increases, it is important to consider the impact of both observational and theoretical systematic errors on these constraints. Here we mention a few of the most important of each.
\subsubsection{Observational systematics}
\label{sec:systematic_obs}
It is worth noting that definitions and measurement uncertainties can matter both qualitatively and quantitatively.  A particularly important systematic error is the fact that in large imaging surveys, the surveys have surface brightness limits and imperfect pipelines that can lead to mis-estimates of the total mass of a galaxy.  For some time, most of the low redshift estimates of the SHMR were based on stellar mass and clustering estimates of SDSS data. These studies may have underestimated the stellar mass/luminosity of brightest cluster galaxies by factors of several at the massive end \citep{kravtsov_etal:14,bernardi_etal:17,huang_etal:17}.  There are several related issues that can impact this estimation.  Differences in the estimation of sky subtraction dominate, but modeling the outer profiles of galaxies, distinguishing between central galaxies and the intracluster light \citep{conroy_etal:07}, and the estimation of stellar masses themselves can also contribute. The high mass end of the SHMR has important consequences for inferences about cooling rates and feedback in massive galaxies, and studies that were based on these earlier results should be re-evaluated in this light.  In particular, based on more recent estimates that indicate increased mass estimates of massive galaxies, group and cluster mass halos should be expected to host larger central galaxies than earlier estimates would have indicated, for example scaling as $\mgal \sim \mhalo^{0.4}$; this can be seen in the estimate of the SHMR in \textbf{Figure 2} from \cite{kravtsov_etal:14} and \cite{behroozi_etal:17} compared to previous estimates.  An additional uncertainty is due to possible changes in the stellar initial mass function at high mass and as a function of radius within a galaxy \citep{bernardi_etal:17, kravtsov_etal:14}.  These systematics can impact the need for effective feedback at the massive end, so they are directly relevant to physical inference from the SMHR.

\subsubsection{Theoretical systematics}

Although gravity is a well-understood process, the results of cosmological $N$-body simulations are still subject to systematic errors that can impact the predictions of models of the galaxy--halo connection. These errors fall into two main categories: resolution and substructure disruption, and robustness of halo finding and tracking.

Lack of spatial and temporal resolution can limit the ability of a simulation to resolve substructure within halos, even with proper mass resolution (see, e.g., \citealt{moore_etal:98,klypin_eta:99,ghigna_etal:00}). \cite{vandenbosch:17} estimates that roughly 80 per cent of all subhalo disruption is numerical rather than physical, 
which may be due to inadequate force softening in simulations \citep{vandenbosch_ogiya:18,vandenbosch_etal:18}. Lack of subhalos equates to lack of satellite galaxies, which can cause models to compare poorly to clustering measurements, or alternatively can drive model selection to parameters that artificially increase the number of satellites. To account for this, some studies incorporate `orphan' subhalos which algorithmically follow the estimated path of disrupted substructure. In some semi-analytic models, the fraction of galaxies represented by orphans can be 10--30\% (\citealt{pujol_etal:17}). To control for this in abundance matching, studies like that by \cite{reddick_etal:13} and \cite{Lehmann17} have compared results from simulations with various resolutions to limit data comparisons to regions in which numerical results are converged, although caution is still warranted given these findings on disruption.

Even if substructure is properly resolved in a simulation, identifying and tracking the substructure robustly can still be a challenge. Many theoretical models require knowledge of the full history of a given halo or subhalo to properly assign its galaxy. Commonly used algorithms such as \textsc{Rockstar} (\citealt{rockstar}) and \textsc{Subfind} (\citealt{springel:01}) do not always yield the same results. Comparisons of different substructure-finding codes, such as \cite{onions_etal:12} and \cite{muldrew_etal11} find generally consistent results, with caveats for when the number of particles in a subhalo becomes small (e.g. less than 50 particles) and when a subhalo passes through the center of a parent halo. In addition to just {\it finding} the subhalos, there is also the need to trace their merger histories. \cite{srisawat_etal:13} compared different merger tree codes, finding distinct results from different codes run on the same simulation. That work proposes a list of necessary features that all codes should contain to produce robust results. 

\section{CURRENT CONSTRAINTS ON SECONDARY PROPERTIES IN THE GALAXY--HALO CONNECTION}
\label{sec:constraints_secondary}

Models of the galaxy--halo connection that parameterize this relationship using $\mhalo$ only have been highly successful in modeling observational data. However, the question remains of just how far the mass-only approach can go to represent the observed galaxy distribution when incorporating galaxy properties other than $\mgal$ and luminosity. The main emphasis of this section is on models of galaxy bimodality: dividing the observed population into blue, star-forming objects and red-and-dead passively evolving galaxies. \S 6.1 reviews current constraints on the SHMR divided along these two lines, while \S 6.2 reviews the pursuit of observational signatures of galaxy assembly bias in galaxy bimodality and galaxy properties other than mass. \S 6.3 discusses how the galaxy--halo connection can explain the observed correlation between bimodality and large-scale environment. Finally, \S 6.4 discusses the correlation between galaxy size and halo mass. 

\begin{figure}
\includegraphics[width=0.9\textwidth]{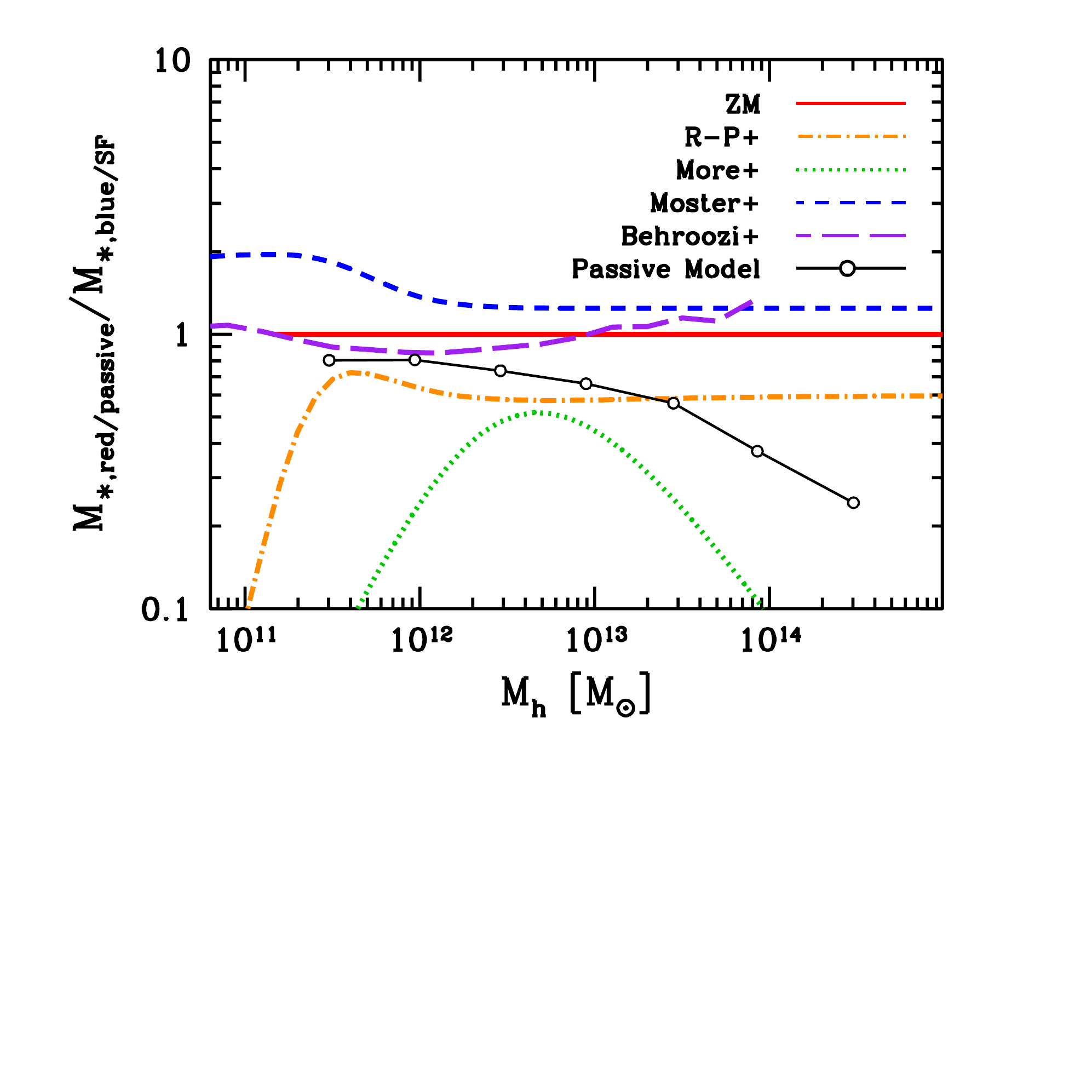}
\caption{The ratio of the SHMR for quenched (red) and star-forming (blue) galaxies at $z=0$ from various analyses in the literature.  All results are for central galaxies only. As the figure demonstrates, there is not yet consensus between approaches. Results here come from \citet{zu_mandelbaum:16}, \citet{rodriguez_puebla_etal:15}, \citet{more_etal:11}, \citet{moster_etal:17}, and \cite{behroozi_etal:17}. See the text for discussion of the various methods involved. The solid black line with open circles is the prediction of the passive quenching model. Each of these studies separately constrain the SHMRs for star-forming and passive galaxies, but here we plot the ratio for brevity.
}
\label{redblue_shmr}
\end{figure}

\subsection{Stellar-to-Halo Mass Relations for star-forming and passive central galaxies}
\label{sec:redblue_shmr}
At present, there is no consensus on the difference in SHMRs between star-forming (blue) and passive (red) central galaxies at $z=0$.  Several examples of estimates of this ratio as a function of halo mass are shown in \textbf{Figure \ref{redblue_shmr}}. 
All of these data are based on the SDSS galaxy sample, but they represent a myriad of analysis techniques, including satellite kinematics \citep{more_etal:11}, empirical abundance modeling \citep{moster_etal:17}, two-point clustering and abundances \citep{rodriguez_puebla_etal:15}, a combination of weak lensing with clustering \citep{zu_mandelbaum:16}, and the forward-modeling Universe Machine of \cite{behroozi_etal:17}. All of these analyses constrain the SHMRs for both subsamples, but for brevity we plot their ratio. These studies find a range of results, including that the quantity $\mred/\mblue$ can be both above or below unity, with one study finding no difference in the SHMR for the two sub-populations. 
Conditional abundance matching models of color make different predictions depending on the exact procedure and which halo properties are chosen to match to. Different models can predict, at fixed $\mhalo$, that passive galaxies can be either more massive, less massive, or the same as star-forming galaxies.

As a simple benchmark for interpretation and comparison to observations, the connected open circles represent a {\it passive quenching model}. In this model, galaxies come off the star-forming sequence randomly and evolve passively, independently of their halo's growth---i.e., the galaxy within a fast-growing $10^{12}$ $\msol$ halo is just as likely to quench as a slow-growing halo of the same mass. Between $z=1$ and $z=0$, the halos grow independently of the state of their central galaxies: the red-sequence galaxies have minimal stellar mass growth, whereas the star-forming galaxies continue to add to their stellar mass. Thus, even if $\shmrr/\shmrb=1$ at $z=1$, by $z=0$ they will be different. The red galaxies will be less massive than the blue galaxies at fixed $\mhalo$. The predictions of this model are also shown in \textbf{Figure \ref{redblue_shmr}}. The indicated trend with mass is expected from the buildup of the red sequence; passive central galaxies at low $\mgal$ have only recently arrived on the red sequence, but massive galaxies quench at very early times, thus there is a large fraction of the age of the universe over which blue galaxies can build up substantially more mass than the quenched galaxies in the same halos. 

To yield a $\mred/\mblue$ ratio different than the passive model, the assumption that galaxy quenching is uncorrelated with halo formation history must break down. Whether or not this breakdown yields galaxy assembly bias in galaxy bimodality---which we have defined as a spatial correlation of galaxy properties (other than mass) at fixed $\mhalo$---is an open question we examine further in this section. Current constraints clearly deviate from the predictions of the passive model. However, given the lack of agreement between $z=0$ studies, we conclude that the $\mred/\mblue$ ratio and its dependence on halo mass remains an open question. 

The differences in these results may come from a myriad of sources. In each approach there are different modeling assumptions, and each uses different statistical quantities to constrain the SHMRs. Each uses different estimates of galaxy stellar mass. \cite{moster_etal:17} and \cite{behroozi_etal:17} use star-formation rate to classify galaxies, whereas the other studies use broadband colors, which can lead to heterogeneous samples. To create progress, each of these assumptions will have to be tested to understand the impact on the SHMR constraints.

One question there is near-consensus on is: Does the SHMR for star-forming galaxies have a pivot point? This question is important because it has consequences for galaxy feedback and the source of the pivot point in the overall SHMR---is the overall pivot point due to more massive galaxies being quenched, or do star-forming galaxies also undergo a transition at $10^{12}$ $\msol$, above which star formation is less efficient even though they are actively star-forming? With the exception of the results from satellite kinematics, which have large error bars at high $\mhalo$, all of the analyses in \textbf{Figure \ref{redblue_shmr}} find that star-forming galaxies do indeed have a pivot point.

The preceding discussion has focused on central galaxies, but satellite galaxies are a major component of the quenched population. Analyses from groups and clusters find that, at all $\mgal$, satellites have a higher quenched fraction than centrals, and that at fixed $\mgal$, the quenched fraction of satellites monotonically increases with $\mhalo$ (see, e.g., \citealt{yang_etal:08,peng_etal:10,wetzel_etal:12_groups1,wetzel_etal:13_groups2,wang_etal:17_elucid4}). The interpretation of this is that the quenching efficiency of satellites is higher than that of central galaxies. Indeed, while there are quenched satellite galaxies at any value of $\mgal$, \cite{geha_etal:12} found that below the limit of $\mgal\approx 10^9$ $\msol$, no isolated central galaxies are quenched. 

\subsection{How does assembly bias manifest observationally for secondary properties?}

If secondary galaxy properties correlate with secondary halo properties that exhibit assembly bias, then this could be detectable in the spatial distribution of galaxies. In this subsection we review results of observational signatures of galaxy assembly bias, focusing on galaxy quenching and galactic conformity. We further compare recent results to predictions from galaxy formation models.

\subsubsection{Galaxy assembly bias and galaxy quenching}

Most emphasis to date regarding observational signatures of galaxy assembly has focused on the galaxy bimodality and whether a galaxy is star-forming or quenched. We demonstrated in the preceding section that studies of the SHMRs for star-forming and passive galaxies are inconclusive. Here we review searches for galaxy assembly bias by directly probing the spatial distribution of galaxies. 

There have been a number of investigations looking for assembly bias in the clustering of red and blue galaxies using group catalogs to estimate halo mass, including \citet{yang_etal:06, wang_etal:08_assembly_bias, lacerna_etal:14}. We regard these results as too susceptible to biases from the group catalog to be reliable. \citet{campbell_etal:15} looked in detail at the results of state-of-the-art group finders and how they assign halo masses to galaxies separated by color. They found that halo masses can be highly biased, possibly explaining differing results in the previous studies. Additionally, \citet{lin_etal:16} used weak lensing to determine unbiased halo mass estimates of galaxy groups---using the \cite{yang_etal:09} group catalog---split into red and blue central galaxies. \citet{lin_etal:16}, focusing on groups at $\mhalo\sim 10^{12}$ $\msol$, found that groups labeled with the same halo mass from the group finder yield different lensing signals, indicating larger halo masses in the groups with red central galaxies. Thus, using the halo masses assigned by the group finder could lead to a spurious detection of assembly bias by yielding a higher clustering amplitude for the groups with red centrals. In contrast, when using lensing measurements to create samples with the same $\mhalo$, the clustering of groups with red central galaxies was the same as the clustering of groups with blue central galaxies. 

\begin{figure}[t]
\includegraphics[width=5in]{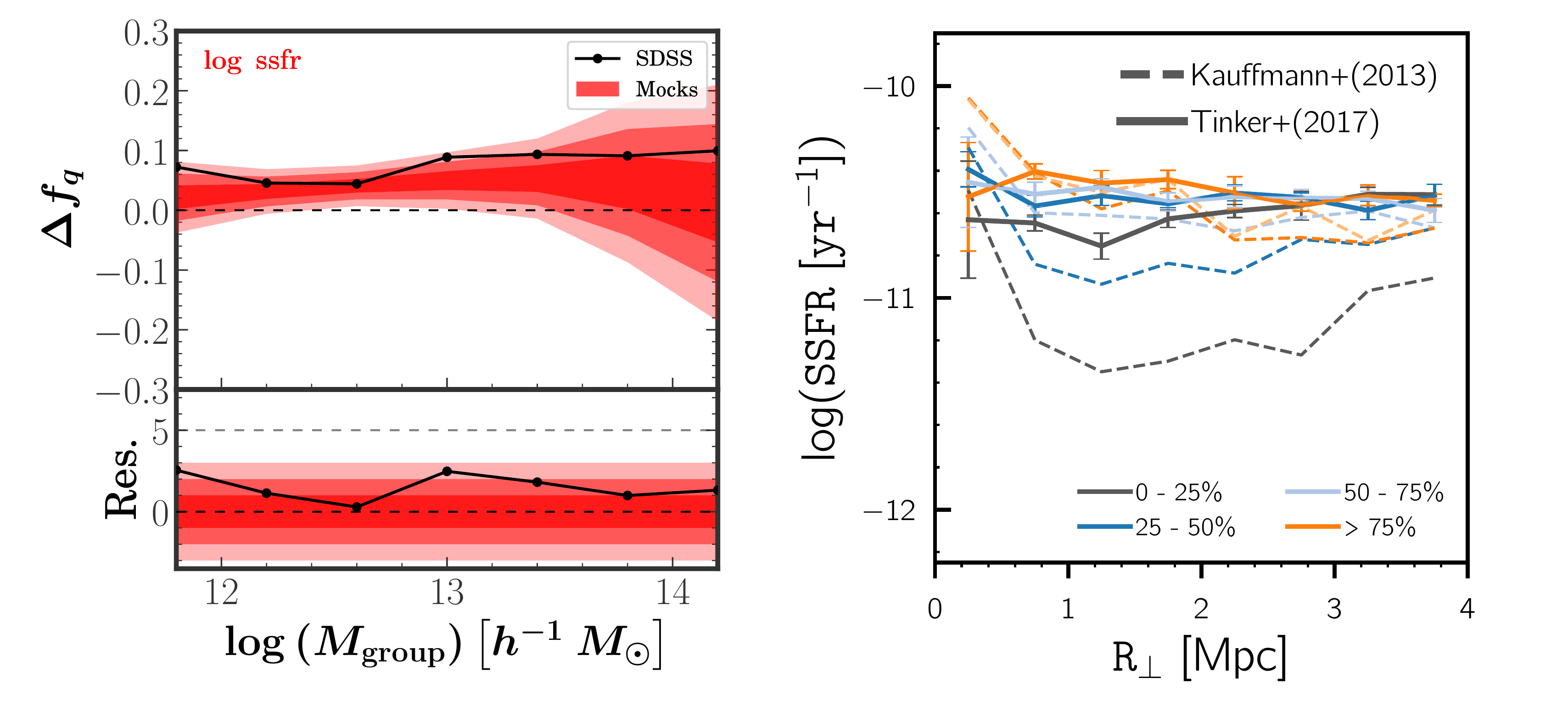}
\caption{{\it Left Panel:} One-halo conformity. The connected black points are measurements of $\Delta f_q$ from SDSS groups, defined as the difference in the quenched fraction of satellite galaxies when the central galaxy is passive and when it is star-forming (see text for further details). $\Delta f_q > 0$ nominally indicates conformity. The shaded regions indicate 1-3$\sigma$ contours from repeating the analysis on mock catalogs with no intrinsic conformity in them. The positive value of $\Delta f_q$ is consistent with that found in mocks, indicating that the conformity seen in the data is from biases in the group catalog. Figure taken from \cite{calderon_etal:17}. {\it Right Panel:} Two halo conformity signal in SDSS galaxies as a function of scale. The curves show the median specific star formation rates (SSFR) for neighbors around central galaxies that are binned by their own SSFR. The bins are defined by their rank order in SSFR as indicated in the panel. The SSFR of the neighbors is measured as a function of projected separation from the central galaxy. The dashed curves show the result of \cite{kauffmann_etal:13}, which was reproduced in \cite{tinker_etal:17_p2}. The solid curves show the result after removing a small number of satellite galaxies that the Kauffmann isolation criteria did not reject. Figure taken from \cite{tinker_etal:17_p2}.}
\label{conformity}
\end{figure}
\subsubsection{Galactic Conformity}

Galactic conformity is the phenomenon that the properties of neighboring galaxies are correlated with---i.e., they `conform to'---the properties of nearby {\it central} galaxies. Conformity studies are usually focused on the color or star formation rates of galaxies. They are divided into two distinct, but possibly correlated, spatial regimes: When the neighboring galaxies of a given central are within the same dark matter halo, this is called one-halo conformity. When the neighboring galaxies are outside the halo virial radius of the central galaxy, this is called two-halo conformity. Conformity can be quantified in a number of ways, but a straightforward test is to look at central galaxies of a given $\mgal$, divided into star-forming and passive samples. The quenched fractions, $\fq$, of neighbors is measured around each of the central subsamples. If $\fq$ for the neighbors around passive centrals is higher than $\fq$ for the neighbors around star-forming centrals, this is a detection of conformity. 

The attention on galactic conformity is driven in part by the possible connection to halo and galaxy assembly bias. For the one-halo term, it is possible to posit quenching mechanisms for central galaxies that also impact the quenching efficiency of satellites. In such a model, the conformity need not be related to assembly bias. But if quenching of central galaxies is related to halo formation history, then it will also be related to the accretion history of satellites---i.e., old parent halos will have old central galaxies as well as old subhalos. Additionally, \cite{hearin_etal:15} proposed two-halo galactic conformity as a sensitive test of galaxy assembly bias. The formation histories of neighboring halos are correlated. Thus, if these formation histories are correlated with quenching mechanisms, two-halo conformity will be present. It is possible to have one-halo conformity without the presence of two-halo conformity, but two-halo conformity by itself necessarily produces a one-halo effect: the satellites in the $z=0$ universe were themselves once neighbors in the field before being accreted.

The first detection of conformity was presented by \cite{weinmann_etal:06}, looking at satellites (the neighbors) around central galaxies in the 1-halo regime using a galaxy group catalog from $z=0$ SDSS data. They found that, at fixed halo mass, the $\fq$ of satellite galaxies was higher when the central was quenched itself, with $\Delta f_q\sim 0.1$--0.2 between passive and star-forming central galaxies. Other analyses have found similar results (e.g., \citealt{knobel_etal:15,kawinwanichakij_etal:16,berti_etal:16}). However, subsequent studies have examined possible systematic biases in the use of the group catalogs to identify halos. \cite{campbell_etal:15} demonstrated that errors in the estimated halo masses can create a false 1-halo conformity signal. \cite{calderon_etal:17} showed that the 1-halo $\Delta f_q$ found in SDSS data was consistent with this false signal, which is shown in left panel of \textbf{Figure \ref{conformity}}. These tests assessing the impact of observational biases do not take into account the possibility that quenched central galaxies live in more massive halos at fixed $\mgal$, which would also induce a 1-halo conformity signal with no galaxy assembly bias, as found in the semi-analytic galaxy model of \cite{wang_white:12}.

The first detection of two-halo conformity was presented by \cite{kauffmann_etal:13}. Their approach to conformity was to identify central galaxies through an isolation criterion, then rank order the central galaxies by their specific star formation rates (sSFRs) and calculated the median sSFR of neighbor galaxies for each quartile in the central sSFR. The measurements showed a substantial decrease in the sSFR of neighbor galaxies around the lowest star-forming central galaxies out to 4 Mpc.  However, further studies by \cite{tinker_etal:17_p2} and \cite{sin_etal:17} have shown this effect to be primarily a selection bias in the isolation criterion. The right-hand panel in \textbf{Figure \ref{conformity}} compares the original \cite{kauffmann_etal:13} result to a new analysis when the small fraction of satellite galaxies are removed from the original sample of `isolated' galaxies. 

Other studies using quenched fractions \citep{berti_etal:16} and marked correlation functions \citep{calderon_etal:17} have found statistically significant two-halo signals, but are small in their actual amplitude, implying $\Delta f_q\sim 1\%$. At this level, it is not clear if the conformity signals found indicates galaxy assembly bias or if they can be explained by various systematic issues (\citealt{tinker_etal:17_p2}). 
Using the {\em conditional abundance matching} framework of \cite{hearin_watson:13}, \cite{tinker_etal:17_p2} compared various implementations of the age-matching model to measurements of two-halo conformity in SDSS galaxies. The low level of two-halo conformity measured in SDSS is inconsistent with a model in which galaxy quenching correlates strongly with halo formation time or similar secondary halo properties. Current data cannot rule out any correlation between these two quantities a $z=0$, but the correlation can only be weak at best. Similar conclusions were reached by \cite{zu_mandelbaum:16}, \cite{zu_mandelbaum:17}, and \cite{tinker_etal:16_p1}.

\subsubsection{Comparison with physical models}

Results from the EAGLE hydrodynamic simulation (\citealt{matthee_etal:17}) and  semi-analytic models (\citealt{croton_etal:07, tojeiro_etal:17}) both show a clear correlation of halo age with total stellar mass of the central galaxy. At fixed $\mhalo$, early-forming halos form more massive galaxies, whereas later-forming halos form less massive ones. This theoretical prediction is in agreement with \cite{zentner_etal:16} and \cite{Lehmann17}, both of which claimed to detect galaxy assembly bias in the total stellar mass of galaxies (under the assumption that more massive galaxies are brighter, which is true for star-forming galaxies but less clear for a mixture of active and passively evolving galaxies). 

Analyzing star-forming central galaxies (i.e., removing all passive galaxies from the sample) from an SDSS group catalog, \cite{tinker_etal:17_p3} found a correlation between SSFR and large-scale density at fixed $\mgal$. These observations were consistent with a conditional abundance matching model in which sSFR was matched to halo growth rate: faster growing halos corresponded to higher-than-average star formers. This agrees with the basic {\em ansatz} of empirical models based on using the time-dependent SHMR to infer galaxy sSFR. Note that these results would not imply large-scale conformity in the quenched fraction of galaxies, since this is only found within the star-forming sub-population. \citet{behroozi_etal:13} and \citet{moster_etal:13} predict that the efficiency of converting accreted baryons into stars for halos with $\mhalo<10^{12}$ $\msol$ peaks at late times. For halos with the same $z=0$ value of $\mhalo$, late-forming halos accrete a higher fraction of their baryons at late times, when this baryon conversion efficiency is maximal. Thus, late-forming halos should have higher stellar masses than early-forming halos, an opposite effect from that seen in the theoretical models. To resolve this situation, further analysis is needed on both the theoretical and observational sides; analyses of clustering using stellar-mass limited samples, marginalizing over cosmological parameters, are required. Predictions of observables from the theoretical models need to be quantified. Observational methods of determining individual halo masses of low-mass field galaxies would be a particularly useful tool.

\begin{figure}[h]
\includegraphics[width=4in]{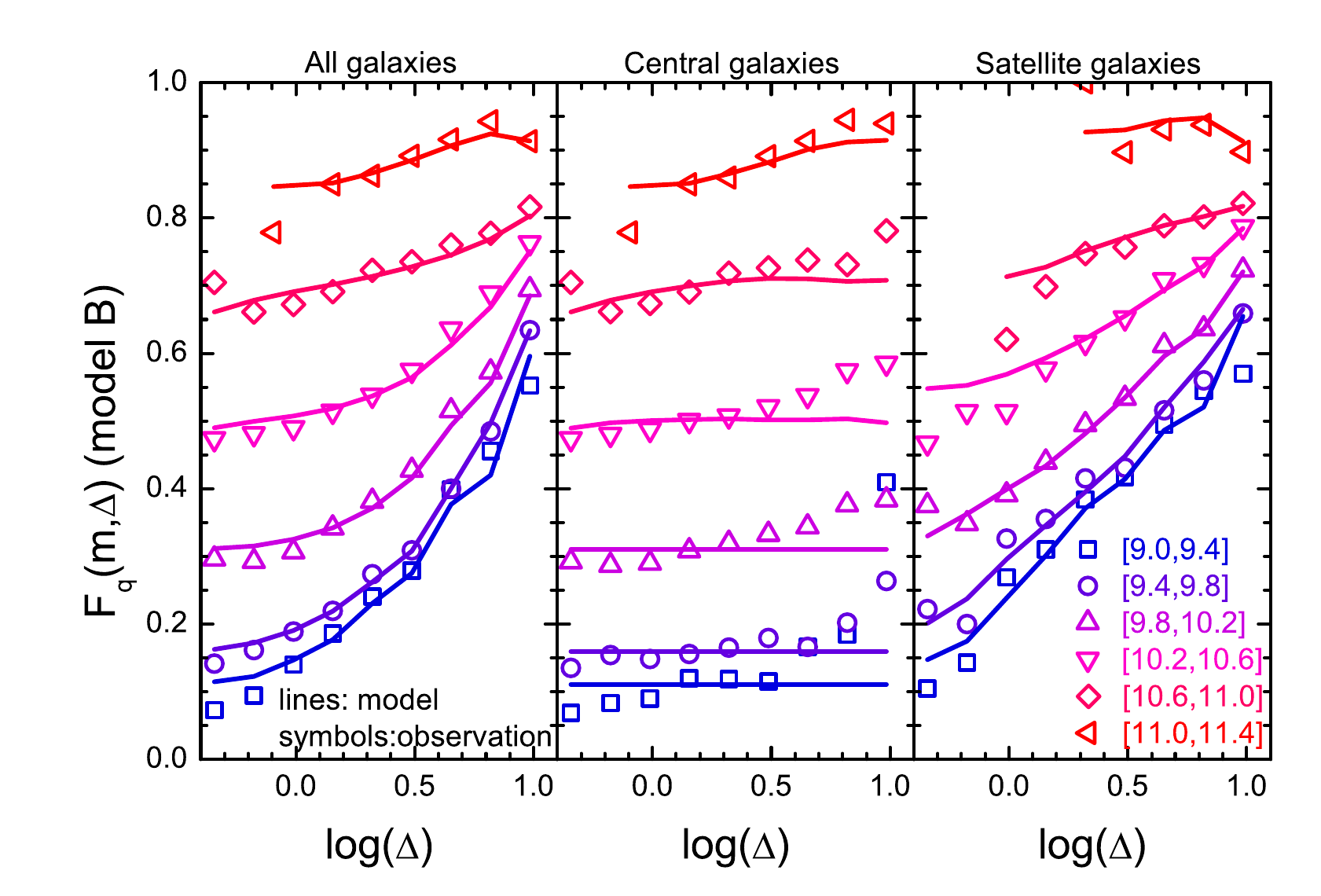}
\caption{The dependence of the galaxy quenched fraction, here expressed as $F_q$, on large-scale density, $\Delta$, measured on $\sim 6$ Mpc scales. The three panels show the relation for all galaxies, central galaxies, and satellite galaxies. Open symbols are measurements from SDSS using the \cite{yang_etal:09} group catalog, whereas solid curves are predictions from a model in which galaxy quenching is only driven by $\mhalo$, and not by $\Delta$ or halo formation history. Adapted from \citet{wang_etal:17_elucid4}.}
\label{elucid}
\end{figure}

\subsection{Why do galaxy properties depend on environment?}
The correlation between galaxy properties and galaxy environment is well established (\citealt{balogh_etal:04,kauffmann_etal:04,blanton_etal:05,blanton_berlind:07}). Properties such as color, star-formation rate, Sersic index, and morphology are each correlated with local galaxy density on all measurable scales. The conclusion of these studies has been that environment mattered most on small scales, and once that environment was fixed, correlations on larger scales were eliminated. However, no `ideal' scale for environment to characterized the full dependence could be identified from the data (\citealt{blanton_berlind:07,blanton_etal:06a}). 

A more comprehensive and physically motivated explanation of the observed correlations of environment can be obtained through the galaxy--halo connection. \textbf{Figure \ref{elucid}}, taken from \cite{wang_etal:17_elucid4}, shows results from the SDSS for the dependence of the quenched fraction of galaxies on galaxy density, $\Delta$, smoothed over a 5.7 Mpc Gaussian kernel. A galaxy group catalog is used to decompose this relation into central and satellite galaxies. The curves show predictions of a model in which $\mgal$ is assigned to $\mhalo$ in a manner similar to abundance matching, but galaxy quenching depends solely on the mass of the host halo. This model is an good match to the observations for all galaxies (left-hand panel), even though $\Delta$ is never explicitly used to determine quenching (see also, \citealt{peng_etal:10,tinker_etal:11}). For low-mass central galaxies, the predictions of the model are by construction nearly horizontal lines, with predictions for higher $\mgal$ bins curved slightly by the trend of $\mhalo$ with $\Delta$ within the $\mgal$ bin, which is stronger at high $\mgal$.  This is an example of spatial correlations in halo mass alone and therefore does not signify assembly bias.
The SDSS measurements for central and satellite galaxies use the group catalog of \cite{yang_etal:07}. For low-mass centrals, the deviation of the data from a horizontal line at high $\Delta$ is due in part to misclassification in the group-finding process \citep{tinker_etal:11,tinker_etal:16_p1}, which may account for the differences with the model predictions. The strongest correlation with $\Delta$ comes from the satellite galaxies. But this correlation, once again, is not due to $\Delta$ {\em per se} but to the change in the halo mass function with $\Delta$; higher mass halos live in higher density regions, yielding the strong correlation of $\fq$ with $\Delta$ for satellite galaxies. 

Intragroup environment---scales of $\sim 100$ kpc or so---also correlates with $\fq$ (\citealt{blanton_berlind:07,peng_etal:10,wetzel_etal:12_groups1}). This correlation can also be explained without invoking environment, through the delayed-then-rapid model of satellite quenching (\citealt{wetzel_etal:13_groups2}). Older satellites are more likely to be quenched, and older satellites are more likely to be near the center of the group, where densities are higher. A model based entirely on time since accretion is consistent with the correlation between $\fq$ and location within the group, although we note that it is not a unique solution to the problem.

\subsection{Galaxy Sizes}
\label{sec:constraints_sizes}
An important question is whether galaxy radial size at fixed halo mass is correlated with a second halo property, and if so what that property is. Classic models of galaxy formation \citep{fall_efstathiou:80,mo_etal:98} have based the radial sizes of galaxies on the spin of their dark matter halos, and predict that they should be proportional to that spin and to the size of the halo.  However, there have been somewhat confusing signals in the literature on whether the {\em amount} of scatter in sizes at fixed galaxy or halo properties are consistent with these models, as well as whether the multivariate correlations between galaxy mass and size at fixed halo mass and radius are what one would predict.

\cite{kravtsov:13} explored the question of how galaxy size is related to halo properties in a standard abundance matching model, matching stellar mass to halo mass. They showed that in this model one obtains an approximately linear relation between the galaxy half-mass radius and the halo radius, with tight scaling across six decades of mass: $R_g \sim 0.015 R_{\rm halo}$, with scatter of 0.2 dex and with fairly similar scaling for elliptical and spiral galaxies.  This agrees well with predictions of simple models where galaxy sizes are set by the angular momentum of their halos.  \citet{somerville:17} compared the conditional size function at a given stellar mass and concluded that the width of the distribution is roughly consistent with the ansatz in which galaxy size is proportional to the standard assumption of spin times the halo size.  However, they found a significant trend in the ratio of galaxy size vs. halo size with both stellar mass and redshift, which is in tension with that ansatz, because the spin should not depend strongly on halo mass or redshift. \cite{desmond_wechsler:17} also found that the resulting scatter in the Tully-Fisher relation was larger than one might expect from these simple models.

The theoretical results of \citet{somerville:17} and \citet{desmond_etal:17}, using different techniques, have found an anti-correlation between stellar mass and size at fixed halo mass, such that smaller galaxies live in more massive halos at fixed $\mgal$. \cite{hearin_etal:17} recently proposed a conditional abundance matching model for galaxy sizes, which came to a similar conclusion based on the clustering properties of small and large galaxies.  This is not by itself conclusive, as galaxy clustering strength is a function of both halo mass and other properties, including satellite fraction and halo assembly bias.  This model is consistent with measuring a higher lensing signal for larger galaxies at fixed stellar mass as was found by \cite{charlton_etal:17}, but we caution that this trend may depend on stellar mass as well as on whether the galaxy samples are color selected.  We expect that upcoming clustering and lensing surveys should be able distinguish these multivariate relations and their evolution with significantly higher precision.
\section{APPLICATIONS OF THE GALAXY--HALO CONNECTION}
\label{sec:applications}
Parameterized descriptions of the galaxy--halo connection provide an effective way to synthesize diverse datasets, and have wide-ranging application in astrophysics and cosmology.  Below, we highlight three of the major areas in which they have been applied: understanding the physics of galaxy formation (\S \ref{sec:galaxyformation}), inferring cosmological parameters (\S \ref{sec:cosmology}), and probing the properties and distribution of dark matter (\S \ref{sec:darkmatter}).

\subsection{Understanding the Physics of Galaxy Formation}
\label{sec:galaxyformation}
We have discussed several of the key insights into galaxy formation that have been informed by studies of the galaxy--halo connection, as well as the interplay between physical and empirical models.  We summarize a few of the most interesting aspects here.

\begin{itemize}
\item \textbf{Which halo properties are most important in setting the properties of galaxies?}
Constraints on the galaxy--halo connection can give us significant and robust information about which properties of dark matter halos and their environments are most important in setting the properties of galaxies.  This includes for example to what extent the star formation rates of galaxies are set by mass, other structural properties of the halos, properties of the mass accretion history, or large scale environment.

\item \textbf{Star formation histories and quenching:}
A new generation of empirical models is now able to trace galaxy histories through time and constrain them with complex combinations of data, including the evolution of the SMF, the relationship between SFR and stellar mass, the evolution of spatial properties with time, and measurements from galaxy lensing.  This has provided significant insight into galaxy star formation histories and quenching timescales over the full range of observed galaxies.  We expect data taken in the near future will provide larger samples to test the spatial and lensing properties at higher redshifts, and provide further insight into the physical mechanisms of star formation and quenching.

\item \textbf{Feedback:} The basic shape of the SHMR has been primary evidence for strong feedback in galaxy formation, over a range of masses (there is additional direct evidence of the processes that lead to this shape, including for example observations of galactic winds).  Although empirical constraints cannot directly probe the physical processes involved, as these constraints improve, they provide increasingly accurate targets for the strength of feedback and its dependence on halo mass, redshift, and environment that any physical model must meet.  Particular examples include that the shape of the SHMR at the massive end constrains the strength of AGN feedback, and the size of the scatter in the SHMR likely puts constraints on what galaxy or halo properties are most responsible for halo quenching.

\item \textbf{Downsizing:} A persistent puzzle in galaxy formation in the context of $\Lambda$CDM was the observation that although CDM predicts that small halos accrete a larger fraction of their dark matter at late times than large halos, small galaxies form a smaller fraction of their stars at late times than large galaxies.  This apparent inconsistency can be understood by the fact that most star formation happens in a fairly narrow band of halo mass (see \textbf{Figure \ref{evolution}} and discussion in \citealt{conroy_wechsler:09} and \citealt{behroozi_etal:13_letter}). A detailed understaning of the combination of physical effects that lead to this narrow range of efficient star formation is still missing.

\item \textbf{Merging, galaxy disruption, and the intracluster light:} Galaxy merging rates have long been considered a test of structure formation and are critical for understanding how galaxies form, but are highly sensitive to the galaxy--halo connection for a given halo population \citep{stewart_etal:09}.
Models with constraints on the galaxy--halo connection over time have made predictions for the buildup of the intracluster light over time and its mass dependence \citep{purcell_etal:07,conroy_etal:07}. Models for the evolution of the galaxy--halo connection can also shed light on what fraction of galaxy build-up is due to mergers; e.g. \cite{behroozi_etal:17} find this to be a strongly increasing function of mass, with nearly all of dwarf galaxy buildup due to {\it in situ} star formation and most of present-day massive galaxy buildup due to mergers.
\end{itemize} 

We expect that as we become more able to empirically constrain the relationship between multi-variate properties of the galaxy--halo connection, constraints on these and other aspects of galaxy formation physics will improve significantly.

\subsection{Inferring Cosmological Parameters}
\label{sec:cosmology}

\begin{figure}[t]
\includegraphics[width=0.95\textwidth]{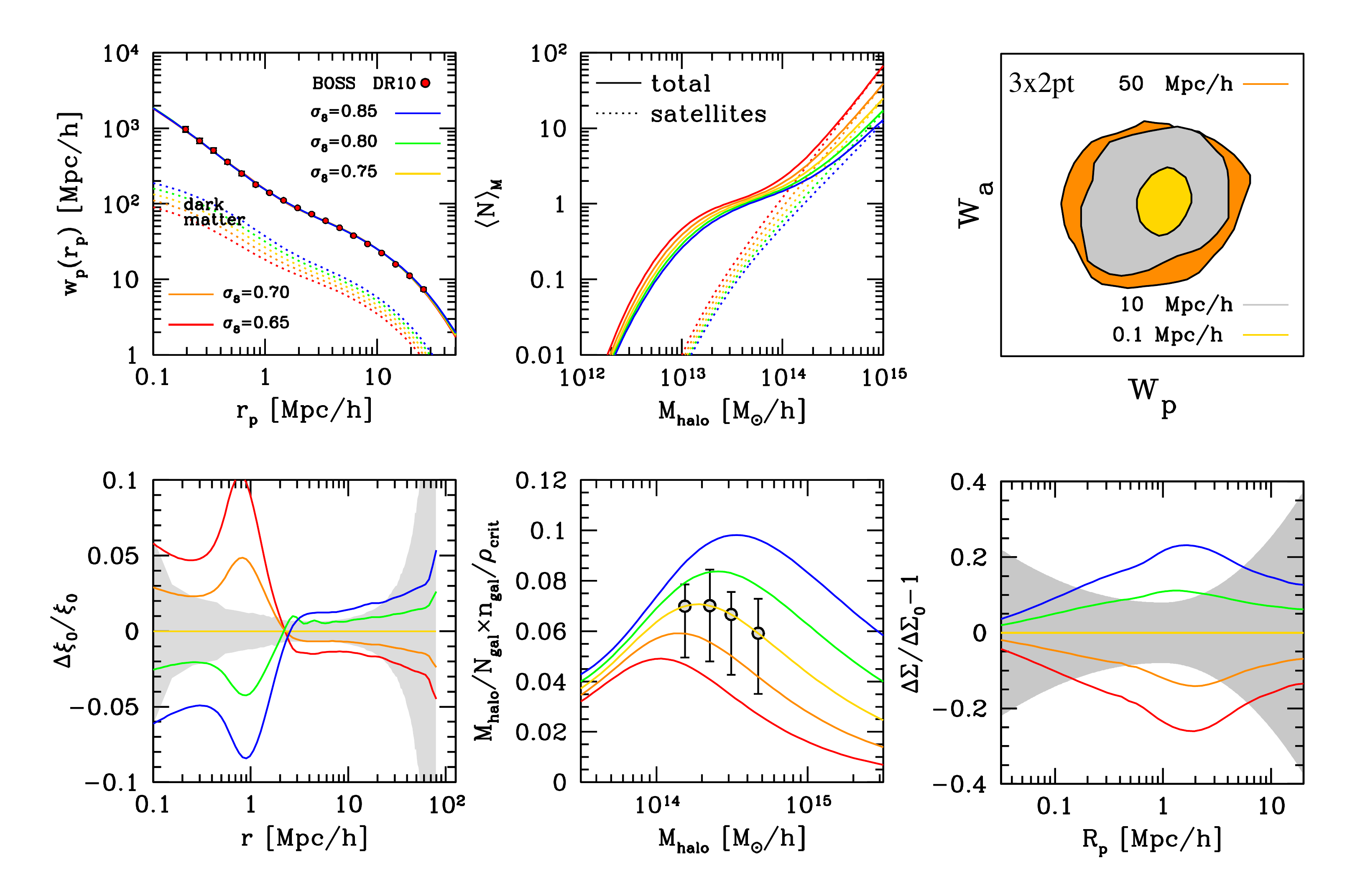}
\caption{{\em Top Left:} Projected galaxy clustering for five cosmological models that vary $\sigma_8$.  Each model is able to match the same two-point galaxy clustering but with different halo occupations.  {\em Top Middle:} Mean halo occupation for the five cosmological models.  Higher values of $\sigma_8$ require fewer galaxies in massive halos, because they have stronger matter clustering.   {\em Top Right:}  Cosmological constraints from the combination of three different two-point correlations (3x2pt): galaxy--galaxy lensing, shear two-point, and angular galaxy--galaxy clustering using an LSST-like photometric survey. The different colors indicate the minimum scale used in the measurements. Bottom panels show three observables that can break the degeneracy between matter clustering and halo occupation.  From left to right, the panels show redshift-space distortions (RSDs), the mass-to-number ratio in clusters, and galaxy--galaxy lensing for each of the values of  $\sigma_8$.  }
\label{cosmo}
\end{figure}

Future galaxy surveys will provide tremendous power for high precision cosmological constraints, especially if clustering measurements can be pushed to smaller scales, within the trans-linear or non-linear regime where the galaxy--halo connection is of increasing importance.  For many statistics, the spatial scale at which the minimum fractional error is achieved is in the range of $1\le r\le 10$ Mpc (see discussion of \textbf{Figure \ref{cosmo}}). This is true even for surveys that are specifically designed to probe structure on linear scales, such as measurements of baryon acoustic oscillations. However, galaxy bias is highly complex at these scales. Higher-order perturbation theory generally breaks down at scales around 10--20 Mpc, or at larger scales for redshift-space clustering (\citealt{carlson_etal:09,wang_etal:14}). Thus, extracting information out of Mpc scales requires a model that is fully non-linear. The galaxy--halo connection provides such a model, provided the model is flexible enough to incorporate any systematic uncertainties, including the accuracy of predictions for scale-dependent halo clustering, galaxy assembly bias, and the impact of baryonic physics on the abundance and clustering of dark matter halos.

Clustering measurements at small scales are sensitive both to growth of structure and to the universal expansion rate, thus providing complementary information to test competing models of cosmic acceleration (see, e.g., the comprehensive review by \citealt{weinberg_etal:13}).  In general, the pathway to cosmological constraints using halo occupation methods starts with measurements of projected galaxy clustering, either through $w_p(r_p)$ or the angular correlation function $\mathsf{}{w}(\theta)$. Using projected quantities is key because they eliminate the effect of redshift-space distortions (RSDs). The top-left panel in \textbf{Figure \ref{cosmo}} shows measurements of the projected clustering of galaxies in the BOSS survey from DR10.  The figure shows halo occupation fits using the analytic model of \citet{tinker_etal:05} with five different values of $\sigma_8$, as listed in the panel (the other cosmological parameters are held fixed).  For reference, the clustering of dark matter is shown for each of these cosmologies. The amplitude of matter clustering, and thus the bias of the galaxies in the halo occupation model, varies significantly with $\sigma_8$, but for each cosmology a good fit can be found to the real-space two-point galaxy clustering. Thus real-space clustering at these scales provides limited information on the amplitude and growth of structure when considered alone. Even though the real-space clustering is the same, each cosmology requires that galaxies occupy different halos (as shown in the top-middle panel).  Thus, the occupation functions constrained by $w_p(r_p)$ can be used to make predictions for statistics that contain more cosmologically sensitive information.  These include RSDs, the mass-to-number ratio of galaxy clusters, and galaxy--galaxy lensing, which are shown in the bottom panels of \textbf{Figure \ref{cosmo}} and discussed in more detail below.

{\bf Redshift-space distortions.} Galaxy redshifts are a function of not just the smooth Hubble flow but also `peculiar' motions caused by the local gravitational potential, which causes galaxies to move toward overdensities and away from underdensities. Thus, the spatial distribution of galaxies in redshift space will have anisotropies due to the amplitude of the velocity field. The velocity field is, in turn, determined by the amount of matter in the universe, how clumpy that distribution is, and by the theory of gravity. Current analyses of RSD yield $\sim 10\%$ measurements of the parameter combination $f\sigma_8$, where $f$ is the logarithmic growth of structure and $\sigma_8$ is the amplitude of matter fluctuations \citep{alam_etal:17}.   The bottom-left panel in \textbf{Figure \ref{cosmo}} shows the variation in the RSD monopole, $\xi_0(r)$, for the five cosmologies,  based on the model by \citet{tinker:07}. This panel also shows the expected measurement error for a full BOSS-like survey based on mock galaxy catalogs.  As discussed above, the `sweet spot' for optimal measurements is in the 1--10 Mpc regime, where sample variance is minimized but shot noise due to small number statistics is avoided. The most constraining power between models comes at $r\sim 1$ Mpc, which represents the transition between pairs of galaxies between two distinct halos and pairs within a single halo. Galaxy pairs within a single halo have larger relative velocities, leading to significant suppression of clustering. As can be seen in the mean occupation functions, the fraction of galaxies that are satellites varies inversely with $\sigma_8$, thus the model with the largest $\fsat$ has the largest pairwise velocity dispersion at $r\sim 1$ Mpc, and the lowest $\xi_0(r)$ at this scale. 

{\bf The mass-to-number ratio of galaxy clusters ($M/N$):} This statistic is analogous to the mass-to-light ratios of galaxy clusters, but reduced the number of free model parameters by simply counting the number of galaxies, $N$, inside a halo. From the mean occupation functions shown in \textbf{Figure \ref{cosmo}}, it is clear that measurements of the mean occupation themselves contain cosmological information. The bottom-middle panel of the figure shows predictions for $M/N$ from the five cosmologies fit to the DR10 BOSS $w_p(r_p)$. Here, the quantity $M/N$ is normalized by the ratio $\rho_{\rm crit}/\bar{n}_{\rm gal}$, where $\rho_{\rm crit}$ is the cosmological critical density, and $\bar{n}_{\rm gal}$ is the mean space density of the galaxies in the sample. Points with errors represent estimates of the uncertainties achievable in a BOSS-like survey at $z<0.3$. Errors in halo masses are taken from the weak lensing analysis of RedMaPPer clusters by \cite{murata_eatl:17}, which are added in quadrature with the expected Poisson noise from the number of clusters in the survey volume (although the mass estimates dominate the error bar). $M/N$ and $M/L$ have been effectively used to constrain cosmological parameters with low-redshift data (\citealt{vandenbosch_etal:03,tinker_etal:12_mn}), and new large-scale redshift and lensing surveys make application to larger volumes imminent. \cite{reddick_etal:14} showed that with current constraining power, simple HODs are sufficient to obtain unbiased parameter constraints, but as the statistical power increases additional parameters may be needed.

{\bf Galaxy--galaxy lensing:} Galaxy--galaxy lensing is a probe of the galaxy--matter cross correlation, and it is sensitive to both the matter density and amplitude of matter fluctuations. The observational quantity measured by galaxy--galaxy lensing is $\DS$, the excess surface mass density at $R_p$, relative to the mean interior density, around a galaxy. The bottom-right panel of \textbf{Figure \ref{cosmo}} shows how $\DS$ varies with $\sigma_8$ for the models fit to the BOSS real-space clustering.  Observational uncertainties in this quantity are shown from \cite{leauthaud_etal:17}, which uses deep CFHTLS (Canada-France-Hawaii Telescope Legacy Survey) imaging in the Stripe 82 field of the BOSS spectroscopic survey. Note that this survey only covers $\sim 200$ deg$^2$, which is only 2\% of the full spectroscopic BOSS catalog.  Even this small area yields constraining power to distinguish these models, indicating the substantial potential of future combinations of large area spectroscopic and imaging surveys. \cite{cacciato_etal:13} and \cite{more_etal:15cosmo} have demonstrated the efficacy of joint clustering and lensing analyses for constraining cosmological parameters in the halo occupation framework.

{\bf 3x2pt:} As discussed above, combinations of two-point statistics can break degeneracies in the galaxy--halo connection and provide powerful cosmological information. A recent study from the Dark Energy Survey \citep{des_2017} used a combination of galaxy--galaxy clustering, shear-shear clustering, and galaxy-shear clustering to put the tightest constraints yet on $\sigma_8$ and $\Omega_m$ in the local Universe (and see related work by \citealt{kilbinger_etal:13} and \citealt{vanuiert_etal:17}).  To date, these analyses have assumed linear bias between the galaxy clustering and matter clustering and exclude small scales where this assumption is expected to fail from the analyses.  However, substantially more constraining power may be available if the modeling can be extended to smaller scales \citep{krause_etal:17}; a full comparison between constraints with a fully nonlinear modeling approach with a parameterized galaxy--halo connection and a quasi-linear approach with a smaller number of bias parameters has yet to be done.

As these examples demonstrate, pushing to smaller scales has significant potential to improve cosmological constraints from current and upcoming datasets, but there are significant challenges to realize this potential, many of which are related to aspects of the galaxy--halo connection.  These include the following:

{\bf Modeling in the non-linear regime:} Historically, researchers have either used fitting functions for the properties of dark matter halos to model galaxy clustering, or have used simulations directly when modeling a range of galaxy clustering models within one cosmological model.  Achieving the required accuracy for these fitting functions is especially challenging in the regime in which there is significant power in the data, 1--10 Mpc.  Methods based on perturbation theory or effective field theory \citep{perko_etal:16} may be effective in the mildly non-linear regime, but they will not be effective in modeling collapsed regions.   The solution may be to emulate the statistics directly (e.g. using techniques similar to those that   \citealt{heitmann_etal:10} used for the dark matter power spectrum), using suites of simulations combined with flexible models of the galaxy--halo connection.

{\bf Assembly bias:} As discussed in Section \ref{sec:assbias}, our understanding of the detailed dependence of galaxy properties on halo properties other than their mass is still in the early stages, and improved modeling will be essential to take small-scale cosmology probes that depend on accurate galaxy clustering models to the next stage. In particular, what is needed is a modeling framework that is flexible enough to encompass the full range of physically plausible manifestations of the complexities of assembly bias for realistic galaxy populations, without losing substantial constraining power; this has yet to be demonstrated.

{\bf Impact of baryons:} Precision cosmology on small-scales will also require understanding the possible range of impacts of galaxy formation and feedback on the matter distribution itself \citep{rudd_etal:08, sembolini_etal:11, schneider_teyssier:15}, including its implications for the mass function and clustering of dark matter halos and subhalos \citep{vandaalen_etal:11, sawala_etal:13, martizzi_etal:14}.  We are still far from being able to simulate the full range of possibilities over a range of cosmological models in order to directly emulate these effects using hydrodynamical simulations, so empirical modeling of the effects, informed by our best physical models, will likely remain the best path forward for the foreseeable future.  The primary impact on the power spectrum can be characterized by a change in galaxy density profiles (\citealt{zentner_etal:08}), but this may not be sufficient for all observable statistics.  Additional observables should be combined to put constraints on the possible amount of feedback; e.g. the small-scale shear power spectrum \citep{foreman_etal:16} and the SZ profiles of groups and clusters \citep{battaglia_etal:17}.

{\bf Intrinsic alignments:} Weak gravitational lensing (see \citealt{mandelbaum:18} for a recent review) depends on the spatial correlations between small distortions in galaxy shapes; if galaxy shapes are aligned with their dark matter halos or with the tidal field, this creates a systematic error that has to be modeled. Different galaxy populations have been shown to have different intrinsic alignments, so in detail one would like to model not just the halo occupation as a function of galaxy properties but also the alignment of galaxies with their halos (\citealt{scheider_bridle:10,blazek_etal:15}).

{\bf Additional aspects:}  Accurate modeling of the galaxy--halo connection will continue to be an important feature of the next generation of cosmological studies, even for those studies that are not pushing to small scales or explicitly including a galaxy--halo connection model.  Examples include the following: 
\begin{itemize}
\item Understanding the error budget in photometric redshift estimates will require effectively modeling the clustering properties of galaxies as a function of their properties \citep{hoyle_etal:17, gatti_etal:17}; this is most effectively done directly through tests with realistic mock catalogs that populate galaxies in halos.
\item Realistic modeling of galaxy clustering on small scales will be required to understand key systematics like fiber selection in spectroscopic surveys and deblending in future imaging surveys \citep{chang_etal:13}.
\item Systematics in cluster cosmology, for example projection and centering effects that impact the mass--richness relationship, can depend on the details of the galaxy--halo connection including its radial distribution and color dependence. 
\end{itemize}

\subsection{Probing the Properties and Distribution of Dark Matter}
\label{sec:darkmatter}
The nature of the dark matter that makes up $\sim 83$\% of the mass in the Universe is still unclear, and an understanding of the galaxy--halo connection can facilitate astrophysical constraints on its nature as well as inform constraints from indirect and direct detection.
Although the $\Lambda$CDM model has had remarkable success on large scales, especially e.g. at larger than the typical sizes of dark matter halos, it is less constrained on smaller scales, where alternative dark matter models can suppress the power spectrum or change the density profiles or dynamics of halos \citep{buckley_peter:17}. 
Understanding and marginalizing over the range of possibilities for the galaxy--halo connection can be critical to robust dark matter constraints in this regime. More generally, there are wide-ranging problems where understanding the properties of the dark matter halos of a specific galaxy or population of galaxies is important, and statistical modeling of the possible halo population of the galaxies provides a way forward. We give a few examples of these applications here.
\begin{itemize}
\item Dwarf galaxies and other probes of small-scale power in CDM are sensitive to the physics of dark matter, but these constraints are in many cases degenerate with uncertainties in the galaxy--halo connection \citep{lovell_etal:12,angulo_etal:13}. There has been significant progress on understanding this interplay in recent years, due to new observations and improved predictions from hydrodynamical simulations, as well as more sophisticated modeling of the galaxy--halo connection. We refer the reader to \cite{bullock_boylankolchin:17} for a more detailed discussion.

\item One of the key tools used to search for WIMP dark matter is indirect detection using gamma rays, looking for the high-energy photons that would be emitted by annihilating dark matter particles in the regions in which they have the highest density \citep[see][for a review]{strigari:13}. The strongest individual sources are the center of the Milky Way itself and its nearby dwarf galaxies \citep{fermi_dwarfs}, but several authors have also considered the stacked signals from groups and clusters of galaxies. For example, \cite{lisanti_etal:17, lisanti_etal:17b} showed that interesting constraints on the dark matter properties could be obtained by looking for excess Fermi signal around hundreds of galaxy groups in the low redshift ($z < 0.03$) Universe.  Accurate mass estimates for the galaxy groups are critical to this estimate, which requires an understanding of the galaxy--halo connection.

\item Is the Milky Way a typical galaxy? Because there are so many measurements that can only be made within the Milky Way or the Local Group, this is a critical question for a wide range of science applications.  Our increasing understanding of the statistical galaxy--halo connection has informed our understanding of the cosmological context of the Milky Way itself.  There is evidence that the Milky Way is more compact than a typical galaxy of its luminosity and circular velocity \citep{licquia_etal:16}, and also that it may have more bright satellites and more quenched classical satellites than a typical halo of its mass \citep{saga,busha_etal:11}.

\item The properties of the Milky Way itself and its relationship to cosmological predictions is also relevant for measurements of direct detection.  In particular, the velocity distribution function of dark matter halos depends significantly on the mass and concentration of the halos; these halo properties are best constrained through the galaxy--halo connection \citep{mao_etal:13, mao_etal:14}.

\item Gravitational time delays in strong lenses provide a measurement of cosmological distance \citep{treu_marshall:16}.  However, additional mass from galaxies and their halos along the line of sight to the systems can also impact the signal and is an important systematic uncertainty in these measurements.  In this case, one has a set of galaxies and would like to know the total mass distribution of the halos surrounding them.  This can be done using models of the galaxy--halo connection as discussed in this review;  \cite{collett_etal:13} showed that knowledge of the external shear could be improved by 30\% using such an approach. 

\item  The predicted amount of substructure in a given system is highly dependent on the mass and concentration of a given dark matter halo \citep{mao_etal:15}.  In order to predict the substructure for a given system of galaxies, one needs a model for the expected mass and concentration given the observed galaxy properties. This is important in modeling strong lensing systems \citep{vegetti_etal:14, hezaveh_etal:16}, as well as for predicting signals from indirect detection. 
\end{itemize}

\section{STATUS AND OUTLOOK} 
\label{sec:status}
Empirical studies of the galaxy--halo connection have exploded since their birth roughly 15 years ago. They are now an essential tool in the interpretation of galaxy surveys for both galaxy formation and cosmology.  These models have provided key insights into the problems of galaxy formation and evolution. They also play an increasing role in cosmological modeling and in understanding the physics of dark matter. 
Accurate methodologies have been developed for modeling the galaxy--halo connection, and there is increasing interplay between modeling approaches (from physical to empirical models, and from simple few parameter models to more flexible models with tens of parameters) that leverages the strengths of each of them.  

Some of the key aspects of the galaxy--halo connection that have been learned from this body of research are as follows:
\begin{itemize}
\item Galaxy formation is surprisingly inefficient: the efficiency of turning gas into stars peaks at $\sim 20$--$30\%$ of the baryon fraction in halos of $10^{12} \msun$ and is significantly lower for higher and lower mass halos.

\item The stellar masses of central galaxies are a strong function of dark matter halo mass below $10^{12} \msun$, scaling like $M_* \propto M_h^{2-3}$, and a weaker function above this pivot point, $M_* \propto M_h^{1/3}$.

\item Galaxy masses are tightly connected to the masses of their dark matter halos.  If halo properties are considered before substantive stripping by more massive systems, the scatter in galaxy stellar mass or luminosity at fixed halo mass has a scatter of less than 0.2 dex for objects above $10^{12}\msun$ where it is well measured; this likely increases by no more than a factor of two at lower mass.

\item For most of the Universe's history from $z\sim$ 8--0, the bulk of all star formation occurs in galaxies that live in a narrow range of halo mass around $10^{12}\msun$.

\item Most galaxies at any stellar mass are the central galaxies in their own dark matter halo; the fraction of satellite galaxies at a given galaxy stellar mass declines from $\sim 30\%$ at low mass to  $\sim 5\%$ at high mass.

\item Most trends of galaxy properties with large-scale environment can be reasonably well explained by the fact that the halo mass function and average halo properties vary with environment, combined with a galaxy--halo connection that is independent of environment.

\item Although halo mass is the dominant determinant of the state of the galaxy occupying it, there is statistically significant evidence that some galaxy properties are influenced by other halo properties, and that this manifests in galaxy clustering properties (assembly bias).
\end{itemize}

The next generation of surveys is likely to transform the study of the galaxy--halo connection into a precision science, including enabling the community to pin down the dependence of the wide range of multi-modal galaxy properties on the key properties of dark matter halos and their environments within the cosmic web.  These surveys include massive imaging and spectroscopic surveys from the ground and space whose combination will jointly constrain the abundance and clustering of galaxies and the mass distribution around them,  as well as surveys of 21-cm, UV absorption, X-rays, and the CMB at high resolution that will  map the gas and its connection to galaxies and their halos.

In the next decade, we expect that our understanding of the detailed connection between galaxies and halos over mass, redshift, and environment will provide major strides forward in galaxy formation, cosmological parameters, and the nature of dark energy and neutrino mass, and in understanding the nature of dark matter.  At the same time, it is clear that for this promise to be realized, the precision of models for the galaxy--halo connection will need to keep up with the pace of the data.  In closing, we highlight some of the most interesting near-term future issues.

\begin{issues}[FUTURE ISSUES]
\begin{enumerate}
\item  The detailed manifestations of assembly bias and their connection to the observable properties of galaxies are still relatively unconstrained. Characterizing them will likely provide interesting insights into galaxy formation physics; in addition, effectively modeling assembly bias will be important to mitigate systematic uncertainties in some cosmological constraints. 
\item The mass dependence of the normalization and scatter in the galaxy--halo connection is still poorly constrained at halo masses below $\sim 10^{12} \msun$. This has important consequences for interpreting measurements of dwarf galaxies in the context of dark matter models.
\item Characterizing the halo occupation of galaxies identified with complex selection criteria, including for example colors, star formation rates, sizes, morphologies, and line widths, will be increasingly important for cosmological studies. 
\item The relationship between galaxy color, galaxy size, and galaxy morphologies and halo properties at fixed stellar mass is still uncertain; there is a need for models and observational tests of models that connect galaxy sizes and morphologies to dark matter halos across cosmic time.
\item Statistically mapping the relationship between halos and the gas surrounding and fueling galaxies is still in the early stages, and constraints on these relationships should provide new physical insight into galaxy formation. 
\item Baryonic processes, especially various forms of feedback, may modify the abundance and clustering properties of dark matter halos, with important implications for inferences about the galaxy--halo connection.
\item A primary challenge for future surveys is optimizing joint constraints on the galaxy--halo connection and cosmological parameters, which will require judicious choices in parameterizing the former to retain maximal constraining power.
\end{enumerate}
\end{issues}

\section*{DISCLOSURE STATEMENT}
The authors are not aware of any affiliations, memberships, funding, or financial holdings that might be perceived as affecting the objectivity of this review. 

\section*{ACKNOWLEDGMENTS}
We are grateful to our many collaborators on these topics over the
years. RHW in particular thanks all of the participants of the 2017
KITP program ``The Galaxy--Halo Connection'' for extensive discussions
that helped frame the perspective of this review, and we thank the
KITP for support via the National Science Foundation under Grant
No. NSF PHY-1125915 while this review was being written.  We thank
Yao-Yuan Mao, Andrew Hearin, and our scientific editor Sandy Faber for
extensive helpful comments on the manuscript.  Susmita Adhikari, Peter
Behroozi, Andreas Berlind, Jonathan Blazek, Joe DeRose, Ashley King, 
Andrey Kravtsov, Sean McLaughlin, Ethan Nadler, Rachel Somerville, Chun-Hao To, and
Frank van den Bosch also provided helpful feedback.  We thank Ralf
Kaehler for assistance with Figures 1 and 6, Sasha Safonova for
assistance with data compilation for Figure 8, and Chang Hahn for
assistance with Figure 11. We thank Peter Behroozi for providing data
for Figures 2, 8, 9, and 10. Figures 1 and 6 made use of the
Chinchilla simulation run at the NERSC supercomputing center.  Figures
3 and 5 made use of the Bolshoi simulations; these were performed
within the Bolshoi project of the University of California
High-Performance AstroComputing Center (UC-HiPACC) and were run at the
NASA Ames Research Center. We have made extensive use of NASA's
Astrophysics Data System and the {\tt arXiv}.

\bibliography{galaxyhalo}
\end{document}